\documentclass[trackchanges, twocolumn]{aastex701}

\newcommand{\pcmsq}{\mbox{$\rm cm^{-2}$}}

\def\ra#1#2#3{#1$^{\rm h}$#2$^{\rm m}$#3$^{\rm s}$}
\def\dec#1#2#3{$#1^\circ #2' #3''$}

\begin{document}

\title{Multiwavelength Observations of GRB\,250419A:\\ A Soft X-ray Flash from a Low-energy Ultrarelativistic Jet}

\author[0000-0001-9915-8147]{Genevieve Schroeder}\affiliation{Department of Astronomy, Cornell University, Ithaca, NY 14850, USA}\affiliation{David A. Dunlap Department of Astronomy and Astrophysics, University of Toronto, 50 St. George Street, Toronto, ON M5S 3H4, Canada}\email{genevieve.schroeder@utoronto.ca}

\author[orcid=0000-0002-9017-3567]{Anna Y. Q. Ho}\affiliation{Department of Astronomy, Cornell University, Ithaca, NY 14850, USA}\email{ayh24@cornell.edu}

\author[0000-0001-5169-4143]{Gavin P. Lamb}\affiliation{Astrophysics Research Institute, Liverpool John Moores University, IC2 Liverpool Science Park, 146 Brownlow Hill, Liverpool, L3 5RF, UK}\email{g.p.lamb@ljmu.ac.uk}

\author[0000-0003-1792-2338]{Tanmoy Laskar}\affiliation{Department of Physics \& Astronomy, University of Utah, Salt Lake City, UT 84112, USA}\email{tanmoy.laskar@utah.edu}

\author[orcid=0000-0001-8472-1996]{Daniel A. Perley}\affiliation{Astrophysics Research Institute, Liverpool John Moores University, IC2 Liverpool Science Park, 146 Brownlow Hill, Liverpool, L3 5RF, UK}\email{D.A.Perley@ljmu.ac.uk}

\author[0000-0003-3115-2456]{Ore~Gottlieb}\affiliation{Department of Physics and Kavli Institute for Astrophysics and Space Research, Massachusetts Institute of Technology, Cambridge, MA 02139, USA}\email{ogot@mit.edu}


\author[0000-0002-2184-6430]{Tom\'as Ahumada}\affiliation{Cerro Tololo Inter-American Observatory/NSF NOIRLab, Casilla 603, La Serena, Chile}\email{tomas.ahumada@noirlab.edu}
\author[0000-0002-8977-1498]{Igor Andreoni}\affiliation{Department of Physics and Astronomy, University of North Carolina at Chapel Hill, Chapel Hill, NC 27599-3255, USA}\email{igor.andreoni@unc.edu}
\author[orcid=0009-0008-2714-2507]{Aleksandra Bochenek}\affiliation{Astrophysics Research Institute, Liverpool John Moores University, IC2 Liverpool Science Park, 146 Brownlow Hill, Liverpool, L3 5RF, UK}\email{A.M.Bochenek@2023.ljmu.ac.uk}
\author[orcid=0000-0001-7511-3745]{Michael Bremer}\affiliation{Institut de Radioastronomie Millimétrique (IRAM), 300 rue de la Piscine, F-38400 Saint-Martin-d’Hères, France}\email{bremer@iram.fr}
\author[0000-0001-5593-0721]{Michael Camilo}\affiliation{Department of Astronomy, Cornell University, Ithaca, NY 14850, USA}\email{mc2923@cornell.edu}
\author[0000-0001-8544-584X]{Jonathan Carney}\affiliation{Department of Physics and Astronomy, University of North Carolina at Chapel Hill, Chapel Hill, NC 27599-3255, USA}\email{jcarney@unc.edu}
\author[0000-0003-1673-970X]{S. Bradley Cenko}\affiliation{Astrophysics Science Division, NASA Goddard Space Flight Center, 8800 Greenbelt Rd, Greenbelt, MD 20771, USA}
\affiliation{Joint Space-Science Institute, University of Maryland, College Park, MD 20742, USA}\email{brad.cenko@nasa.gov}
\author[0000-0001-8104-3536]{Alessandra~Corsi}\affiliation{William H. Miller III Department of Physics and Astronomy, Johns Hopkins University, Baltimore, Maryland 21218, USA}\email{acorsi2@jh.edu}
\author[orcid=0000-0002-8262-2924]{Michael W. Coughlin}\affiliation{School of Physics and Astronomy, University of Minnesota, Minneapolis, MN 55455, USA}\email{cough052@umn.edu}
\author[0009-0006-7990-0547]{James~Freeburn}\affiliation{Centre for Astrophysics and Supercomputing, Swinburne University of Technology, Victoria 3122, Australia}
\affiliation{ARC Centre of Excellence for Gravitational Wave Discovery (OzGrav), Victoria 3122, Australia}
\affiliation{Department of Physics and Astronomy, University of North Carolina at Chapel Hill, Chapel Hill, NC 27599-3255, USA}\email{jfreeburn@swin.edu.au}

\author[0000-0002-5698-8703]{Erica Hammerstein}\affiliation{Department of Astronomy, University of California, Berkeley, CA 94720-3411, USA}
\affiliation{Berkeley Center for Multi-messenger Research on Astrophysical Transients and Outreach (Multi-RAPTOR), University of California, Berkeley, CA 94720-3411, USA}\email{ekhammer@berkeley.edu}
\author[orcid=0000-0002-7778-3117]{Rahul Jayaraman}\affiliation{Department of Astronomy, Cornell University, Ithaca, NY 14850, USA}\email{rj438@cornell.edu}
\author[0000-0002-5619-4938]{Mansi M. Kasliwal}\affiliation{Division of Physics, Mathematics and Astronomy, California Institute of Technology, Pasadena, CA 91125, USA}\email{mansi@astro.caltech.edu}
\author[0000-0002-7465-0941]{Noel Klingler}\affiliation{Center for Space Sciences and Technology, University of Maryland, Baltimore County, Baltimore, MD, 21250, USA} \affiliation{Astroparticle Physics Laboratory, Astrophysics Science Division, NASA Goddard Space Flight Center, Greenbelt, MD 20771, USA} \affiliation{Center for Research and Exploration in Space Science and Technology, NASA Goddard Space Flight Center, Greenbelt, MD 20771, USA}\email{noelklin@umbc.edu}
\author[0000-0002-9700-0036]{Brendan O'Connor}\affiliation{Department of Astronomy, University of Maryland, College Park, MD 20742, USA}
\affiliation{Joint Space-Science Institute, University of Maryland, College Park, MD 20742, USA}
\affiliation{Astrophysics Science Division, NASA Goddard Space Flight Center, 8800 Greenbelt Rd, Greenbelt, MD 20771, USA}\email{oconnorb@umd.edu}
\author[orcid=0009-0003-2780-704X]{Cassie Sevilla}\affiliation{Department of Astronomy, Cornell University, Ithaca, NY 14850, USA}\email{jms949@cornell.edu}
\author[0000-0002-6428-2700]{Gokul P. Srinivasaragavan}\affiliation{Center for Interdisciplinary Exploration and Research in Astrophysics (CIERA), Northwestern University, 1800 Sherman Ave., Evanston, IL 60201, USA}\email{gokul.srinivasaragavanq@northwestern.edu}
\author[0000-0003-2434-0387]{Robert Stein}\affiliation{Department of Astronomy, University of Maryland, College Park, MD 20742, USA}
\affiliation{Joint Space-Science Institute, University of Maryland, College Park, MD 20742, USA}
\affiliation{Astrophysics Science Division, NASA Goddard Space Flight Center, 8800 Greenbelt Rd, Greenbelt, MD 20771, USA}\email{rdstein@umd.edu}
\author[0009-0000-4044-8799]{Jada L.~Vail}\affiliation{Department of Astronomy, Cornell University, Ithaca, NY 14850, USA}\email{jlv93@g.cornell.edu}
\author[0000-0002-3158-6820]{Sylvain Veilleux}\affiliation{Department of Astronomy, University of Maryland, College Park, MD 20742, USA}
\affiliation{Joint Space-Science Institute, University of Maryland, College Park, MD 20742, USA}\email{veilleux@umd.edu}
\author[0000-0003-1710-9339]{Lin Yan}\affil{Caltech Optical Observatories, California Institute of Technology, Pasadena, CA 91125, USA}\email{lyan@caltech.edu}

\begin{abstract}

Soft X-ray flashes (XRFs) are predicted from a variety of hypothesized phenomena, including ``dirty fireballs'' (relativistic jets with significant baryon loading), and gamma-ray bursts viewed marginally off-axis. For a given burst, detailed multiwavelength observations are necessary for distinguishing between possible progenitor scenarios, but the number of XRFs with such data remains small. Here we present X-ray, optical, and radio (mm to cm) data of GRB\,250419A, an XRF (with peak energy $E_p<4\,$keV) discovered by the Space-based multi-band astronomical Variable Objects Monitor. We observe a clear achromatic break in the X-ray and optical bands, and likely two spectral components in the radio emission. The afterglow is similar to the long-duration gamma-ray burst population, but slightly underluminous, and the optical afterglow has an unusually shallow decay phase at $\lesssim2\,$d. We find that the event is well explained by a high-Lorentz factor jet with low kinetic energy ($E\sim10^{49}$--$10^{50}\,$erg), with energy injection that could arise from radial stratification and/or the jet being viewed marginally off-axis. We conclude that at least some X-ray flashes are an extension of classical gamma-ray bursts down to lower energies. 

\end{abstract}


\section{Introduction}
\label{sec:Introduction}

Long duration $\gamma$-ray bursts (GRBs) are cosmological explosions with isotropic-equivalent energies of $\sim 10^{49}$--$10^{54}~$erg \citep{2004RvMP...76.1143P, 2015PhR...561....1K}. Their association with star-forming host galaxies and stripped-envelope supernovae have solidified their connection to the final moments of a massive star's life \citep{2006ARA&A..44..507W, 2010MNRAS.405...57S, 2011AN....332..434M, 2012grb..book..169H, 2013ApJ...778..128P, 2016ApJ...832..108M, 2016SSRv..202..111P, 2017AdAst2017E...5C}. The current physical model of a GRB is that it is the result of a newly born central engine producing a relativistic collimated jet. As this jet (with half opening angle $\theta_{\rm j}$) interacts with the surrounding circumburst material, it accelerates electrons, which in turn produce non-thermal synchrotron emission known as the ``afterglow'' \citep{1998ApJ...497L..17S, 1999ApJ...523..177W, 1999ApJ...519L..17S, 2002ApJ...568..820G}. The broad-band (X-ray to radio) behavior of the afterglow is not only dependent on the energetics and microphysics of the explosion and the environment of the progenitor, but also the jet structure and the viewing angle of the observer \citep[$\theta_{\rm v}$, ][]{2002ApJ...570L..61G, 2003ApJ...591.1075K, 2010ApJ...722..235V}. Through monitoring and modeling of the broad-band afterglow these various physical and geometric properties can typically be constrained.

Two natural expectations of the GRB model are that: 1) some GRB jets must be directed away from Earth (and thus viewed off-axis) and 2) some jets must have a higher baryon fraction, resulting in a ``dirty fireball'' with a lower initial Lorentz factor ($\Gamma_0 \sim 10$ instead of $\Gamma_0 > 100$) \citep{1998ApJ...494L..45P, 1999ApJ...513..656D}. Several avenues of identifying off-axis GRBs and dirty fireballs have been attempted, including searching for fast optical transients \citep[e.g.][]{2013ApJ...779...18B, 2008ApJ...682.1205R, 2013ApJ...769..130C, 2022ApJ...938...85H, 2025MNRAS.537.2362P} and searching for late-rising radio afterglows following stripped-envelope supernovae \citep[e.g. ][]{2003ApJ...599..408B, 2006ApJ...638..930S, 2016ApJ...830...42C, 2019ApJ...879...89M, 2023ApJ...953..179C, 2025ApJ...995...61S, 2026ApJ..1002..194O}. However, these approaches have limitations---for optically discovered events it is often difficult to rule out a classical GRB counterpart (e.g., \citealt{2020ApJ...905...98H}), and poor constraints on the burst time lead to modeling degeneracies (e.g., \citealt{2025ApJ...985..124L}). A promising avenue for resolving these issues is the identification of so-called ``X-ray flashes" (XRFs), which have been theorized to originate from off-axis GRBs and dirty fireballs. 

XRFs are an observational class of cosmological events which are characterized by prompt emission detected by high energy (e.g. X-ray/gamma-ray) satellites with the peak of their prompt emission falling below $\sim 10~$keV. Early X-ray satellites \citep[e.g. BeppoSAX, MAXI, HETE-2][]{2001grba.conf...16H, 2005ApJ...629..311S, 2016PASJ...68S...1N, 2026arXiv260600225P} detected XRFs, but their characterization remained difficult due to limited multiwavelength follow-up observations. While the launch of NASA’s Neil Gehrels \emph{Swift} Observatory and \emph{Fermi} \citep{2004ApJ...611.1005G, 2009ApJ...702..791M} in 2004 and 2008 (respectively) led to a significant increase in the detection, localization, and broad-band follow-up of GRBs, the energy ranges of the detectors on these satellites ($> 15$~keV for \emph{Swift} and $>8~$keV for \emph{Fermi}) preclude them from efficiently detecting XRFs. However, the 2024 launch of two sensitive X-ray survey facilities---the Einstein Probe \citep[EP; ][]{2018SPIE10699E..25Y, 2022hxga.book...86Y, 2025SCPMA..6839501Y} equipped with the Wide-field X-ray Telescope ($0.5$--$4~$keV) and the Space-based multi-band astronomical Variable Objects Monitor \citep[SVOM; ][]{2016arXiv161006892W} equipped with ECLAIRs ($4$--$150~$keV)---mean that XRFs are now being routinely detected
(e.g., \citealt{2025NatAs...9.1073S, 2025arXiv250417034L, 2025arXiv251210239S, 2026arXiv260326213D, 2025ApJ...988L..34J, 2026arXiv260610011R, 2026arXiv260420346S, 2026arXiv260701090D}).

In the off-axis scenario, an XRF is expected when the jet is viewed only slightly off-axis ($\theta_{\rm v} \sim \theta_{\rm j}$). The off-axis view results in lower Doppler boosting and higher pair production optical depth, suppressing the gamma-ray flux relative to the X-rays. Alternatively, X-rays can be produced by the shocked cocoon of the relativistic jet \citep{2002ApJ...571L..31Y, 2002MNRAS.337.1349R, 2004ApJ...608..365Z, 2005ApJ...630.1003G}. In the dirty fireball scenario, the XRF is the result of the slower moving jet producing lower energy emission when viewed on-axis \citep{1999ApJ...513..656D, 2002MNRAS.332..735H}. Additionally, an XRF may be the result of an intrinsically low energy on-axis GRB \citep{2025ApJ...988L..34J, 2026arXiv260420346S}. The advantage of identifying dirty fireballs and off-axis GRBs through XRFs is that the time of first light is precisely known, and the properties of the high-energy prompt emission can be used to help discriminate between progenitor models.

Here we present the broad-band observations and analysis of GRB\,250419A, an XRF detected by SVOM. In Section~\ref{sec:Observations} we discuss the prompt emission and present X-ray, optical, and radio follow-up of the long-lived afterglow. In Section~\ref{sec:Analysis} we interpret the observed emission in the context of synchrotron afterglow emission and compare GRB\,250419A to other long GRBs and XRFs. We model the emission within the context of a jetted synchrotron afterglow in Section~\ref{sec:afterglowmodel}. We explore the implications of our results in Section~\ref{sec:Discussion} and conclude in Section~\ref{sec:Conclusion}. We use the convention $F_{\nu} \propto t^{\alpha}\nu^{\beta}$ and employ the $\Lambda$CDM cosmological parameters of $H_{0} = 68~{\rm km \, s}^{-1} \, {\rm Mpc}^{-1}$, $\Omega_{M} = 0.31$, $\Omega_{\rm \Lambda} = 0.69$ \citep{2020A&A...641A...6P}.

\section{Observations}
\label{sec:Observations}

\subsection{SVOM Detection}
\label{sec:svom_detection}

GRB\,250419A triggered the ECLAIRs instrument onboard SVOM on 19 April 2025 at 02:29:32 UTC and was localized to R.A.(J2000) =  \ra{13}{29}{32.97}, Dec. (J2000) \dec{07}{05}{09.92} with a positional uncertainty of 9.15\arcmin\ \citep[90\% confidence, ][]{2025GCN.40168....1W}. According to initial analysis of the SVOM/ECLAIRs data, the event is double peaked with a precursor pulse with duration $\sim 15~{\rm s}$ occurring $\sim 120~{\rm s}$ prior to the SVOM/ECLAIRs trigger. The duration of the second pulse is measured to be $T_{90}({\rm 5\mbox{--}8 keV}) = 11.44_{-2.58}^{+2.85}~{\rm s}$. The peak energy of the event is $E_{\rm peak} < 4~{\rm keV}$, classifying GRB\,250419A as an XRF \citep[][]{2025GCN.40234....1B}. 

The time averaged spectra of the two pulses can be fit either with a simple power law (pl) or a black body (bb) model. For the pl model, the emission can be fit with photon indices of $3.01_{-0.69}^{+0.80}$ (first pulse) and $3.02_{-0.53}^{+0.56}$ (second pulse), with a total fluence of $f_{\gamma, \rm pl} ({\rm 4\mbox{--}10 keV}) = 1.41_{-0.03}^{+0.03} \times 10^{-7}~{\rm erg~cm}^{-2}$ in the 4--10 keV band. For the bb model, the first pulse has a redshift-corrected temperature of $2.44_{-0.54
}^{+0.64}~$keV and the second pulse has a redshift-corrected temperature of $2.28_{- 0.31}^{+0.38}~$keV, with a total fluence of $f_{\gamma, \rm bb} ({\rm 4\mbox{--}10 keV}) = 1.38_{-0.10}^{+0.06} \times 10^{-7}~{\rm erg~cm}^{-2}$ in the 4--10 keV band \citep[][]{2025GCN.40234....1B}. We integrate the two models over 1--10000~keV (the standard rest-frame range used to compare events across multiple detection facilities) to calculate the bolometric correction factor ($k_{\rm bol}$), and find $k_{\rm bol, pl} \approx 5.6$ and $k_{\rm bol, bb} \approx 1.9$.

\subsection{Targeted X-ray Follow-up}
\begin{deluxetable}{ccccccccc}
\tablecaption{X-ray observations of GRB\,250419A \label{tab:xray}}
\tablehead{
\colhead{Start Time} & \colhead{$\delta t$} & \colhead{Count Rate} &
\colhead{Flux} & \colhead{Instrument} \\
\colhead{(MJD)} & \colhead{(d)} & \colhead{($10^{-2}\,$ct/s)} &
\colhead{($10^{-13}$\,erg/cm$^2$/s)}
}
\startdata
60784.18087 & 0.08 & $11.07^{+0.95}_{-0.95}$ & $40.89^{+3.52}_{-3.52}$ & Swift/XRT\\
60784.70530 & 0.60 & $2.52^{+0.39}_{-0.39}$ & $9.32^{+1.43}_{-1.43}$ & Swift/XRT\\
60787.44345 & 3.34 & $0.43^{+0.22}_{-0.17}$ & $1.58^{+0.82}_{-0.62}$ & Swift/XRT\\
60788.62507 & 4.52 & $<0.55$ & $<2.02$ & Swift/XRT\\
60791.20896 & 7.11 & $<0.43$ & $<1.57$ & Swift/XRT\\
60800.63675 & 16.53 & $<0.008$ & $<0.03$ & Chandra/ACIS\\
\enddata
\tablecomments{All count rates and fluxes are in the 0.3--10 keV band. Upper limits are $3\sigma$.}
\end{deluxetable}

\subsubsection{Swift/XRT}

Five observations were obtained of GRB\,250419A with the X-Ray Telescope (XRT; \citealt{Burrows2005}) on board
the Neil Gehrels \emph{Swift} Observatory \citep{2004ApJ...611.1005G}, including under a target-of-opportunity program (PI: Gompertz). Observations spanned $\delta t=0.08$--7.1\,d (where $\delta t$ is the time since the SVOM/ECLAIRs trigger). We reduced the XRT data using the online tool\footnote{\url{https://www.swift.ac.uk/user_objects/}} from the \emph{Swift} team \citep{Evans2007,Evans2009}, with centroiding turned on and the data binned by observation. The source was detected in the first three epochs (until $\delta t=3.3\,$d) and was not detected in the final two epochs. 
To convert count rate to flux,
we used a hydrogen column density $n_H = 2.25 \times 10^{20}\pcmsq$ \citep{Willingale2013} and
a photon power-law index of $\Gamma=2$.
The resulting values are shown in Table~\ref{tab:xray}.

\subsubsection{Chandra}

We obtained one 20\,ks observations with the \emph{Chandra} X-ray Observatory under Director's Discretionary Time\footnote{PI Ho, ObsId 30917, \dataset [ADS/Sa.CXO\#obs/23171]} ($\delta t=16.5~$d). The data were reduced using
the Chandra Interactive Analysis of Observations (CIAO) software package (v4.12; \citealt{Fruscione2006}).
We used a 2$^{\prime\prime}$-radius region centered on the source.
To measure the background, we set an annulus around the source with an inner radius of 4$^{\prime\prime}$ and an outer radius of 11$^{\prime\prime}$. 
We found that there was no detection, with an upper limit on the 0.5--8\,keV count rate of $1.2\times10^{-4}$ (90\% confidence). 
We used WebPIMMS to convert the upper limit on the count rate to an upper limit on the 0.3--10\,keV flux,
assuming the same photon index and $n_H$ value as for the \emph{Swift} observations. The results are listed in Table~\ref{tab:xray}. 

\subsection{Optical}
\label{sec:optical_followup}

The optical afterglow of GRB\,250419A was observed and detected by several optical, infrared, and ultraviolet telescopes \citep[first discovered by the COLIBR\'I telescope at $\delta t \approx 0.6~$hr;][]{2025GCN.40169....1O}. Spectroscopic observations of the optical afterglow of GRB\,250419A led to a redshift determination of $z = 0.845$ \citep[][]{2025GCN.40174....1T}. Here, we describe our optical observations of GRB\,250419A, which were coordinated through the Fritz SkyPortal Marshal \citep{2019JOSS....4.1247V, 2023ApJS..267...31C}. A summary of our observations can be found in Table~\ref{tab:opticalobs}. We additionally include optical observations from the General Coordinates Network (GCN) in our subsequent analysis \citep{2025GCN.40169....1O, 2025GCN.40170....1X, 2025GCN.40171....1Z, 2025GCN.40172....1K, 2025GCN.40179....1L, 2025GCN.40180....1O, 2025GCN.40182....1P, 2025GCN.40183....1P, 2025GCN.40185....1K, 2025GCN.40186....1W, 2025GCN.40187....1X, 2025GCN.40188....1J, 2025GCN.40189....1G, 2025GCN.40202....1P, 2025GCN.40209....1C, 2025GCN.40210....1L, 2025GCN.40238....1D, 2025GCN.40245....1P}.

\startlongtable
\begin{deluxetable*}{cccc}
\tablecolumns{4}
\tablecaption{Optical observations of GRB\,250419A 
\label{tab:opticalobs}}
\tablehead{\colhead{Observatory} & \colhead{$\delta t^{a}$} & \colhead{Band}  & \colhead{Magnitude$^{b}$}
\\ 
\colhead{} & \colhead{(d)} & \colhead{} & \colhead{AB (mag)}
}
\startdata
LT/IO:O & 0.814 & $r$ & $19.96 \pm 0.06$\\
 & 0.823 & $i$ & $19.76 \pm 0.06$\\
 & 1.805 & $r$ & $20.61 \pm 0.09$\\
 & 1.814 & $i$ & $20.41 \pm 0.08$\\
 & 2.802 & $i$ & $21.29 \pm 0.08$\\
 & 2.81 & $r$ & $21.65 \pm 0.1$\\
 & 3.798 & $r$ & $22.20 \pm 0.15$\\
 & 3.811 & $i$ & $22.06 \pm 0.14$\\
 & 4.878 & $i$ & $22.91 \pm 0.22$\\
 & 5.831 & $i$ & $23.53 \pm 0.31$\\
 \hline
LDT/LMI & 6.092 & $r$ & $23.60 \pm 0.15$\\
 & 6.101 & $i$ & $23.43 \pm 0.17$\\
 & 8.26 & $r$ & $24.11 \pm 0.16$\\
 & 8.271 & $i$ & $24.35 \pm 0.30$\\
 & 10.187 & $r$ & $24.73 \pm 0.50$\\
 & 10.194 & $i$ & $24.38 \pm 0.19$\\
 \hline
SOAR/GHTS & 7.101 & $r$ & $23.81 \pm 0.34$\\
 & 7.113 & $i$ & $23.54 \pm 0.25$\\
 & 8.976 & $r$ & $23.25 \pm 0.50$\\
 & 8.988 & $i$ & $22.96 \pm 0.50$\\
\hline
\enddata
\tablenotetext{a}{Midtime of observation compared to SVOM 
trigger, observer frame}
\tablenotetext{b}{Magnitudes are not corrected for Galactic extinction.}
\end{deluxetable*}

\subsubsection{Liverpool Telescope}

We observed the position of GRB\,250419A with the Liverpool Telescope \citep[LT; ][]{Steele2004} using the IO:O camera in the SDSS $r$ and $i$-bands at multiple epochs, starting at $\delta t \approx 0.81~$d after SVOM trigger. Basic reductions (bias subtraction, flat fielding, and astrometry) of each frame were provided by the automatic IO:O pipeline. A small number of images in the fourth and sixth epochs were affected by tracking errors or poor seeing, and were consequently discarded. $r$-band images from the third epoch were affected by incorrect astrometry and corrected with the use of \texttt{Astrometry.net} \citep{Lang2010}. Remaining frames were stacked with the aid of \texttt{SWarp} software \citep{SWarp}, and PSF photometry on the obtained stacked images was performed with the help of \texttt{SExtractor} \citep{SourceExtractor} and \texttt{PSFex} \citep{Bertin2013}, using the Panoramic Survey Telescope and Rapid Response System 1 (Pan-STARRS1) catalogue \citep{Chambers2016} as reference.

\subsubsection{Lowell Discovery Telescope}
We observed the position of GRB\,250419A with the Large Monolithic Imager (LMI) on the Lowell Discovery Telescope (LDT) on three epochs, between 2025-04-25 and 2025-04-29 ($\delta t \approx 6.1$--$10.2~$d) in $r$ and $i$-bands.  We reduced the images using a custom \texttt{Python}-based image analysis pipeline \citep{Toy2016}. This pipeline performs data reduction, astrometry, registration, source extraction and PSF photometry using \texttt{SourceExtractor} \citep{SourceExtractor}. \texttt{SourceExtractor} was calibrated using point-sources from PanStarrs1 (PS1) Data Release 1 catalog \citep{Chambers2016}. 

\subsubsection{Southern Astrophysical Research Telescope}

We imaged the location of GRB\,250419A with the Goodman High Throughput Spectrograph (GHTS) mounted on the Southern Astrophysical Research Telescope (SOAR) in $r$ and $i$-bands across two epochs, $\delta t \approx 7.1,~ 9$.  Three 300\,s exposures each per filter were obtained for both epochs.  We perform bias subtraction, flat fielding, background subtraction, astrometry with \texttt{Astrometry.net} \citep{Lang2010} and stack the images using \texttt{SWarp} \citep{SWarp}.  Aperture photometry was obtained from resultant, stacked images and calibrated using the Panoramic Survey Telescope and Rapid Response System 1 (Pan-STARRS1) catalogue \citep{Chambers2016}.

\subsection{Radio}

\subsubsection{VLA Observations}

We observed the position of GRB\,250419A with the NSF's Karl G. Jansky Very Large Array (VLA) at a mid-time of $\delta t \approx 3.2~$days (Program 25A-374, PI Perley) at a central frequency of 10~GHz ($X$-band, bandwidth of 4~GHz). We used 3C286 for flux and bandpass calibration and J1309+1154 for complex phase calibration. We reduced the data using the Common Astronomy Software Applications \citep[\texttt{CASA}][]{2007ASPC..376..127M, 2022PASP..134k4501C} VLA Calibration Pipeline\footnote{\url{https://science.nrao.edu/facilities/vla/data-processing/pipeline/}} and imaged the data using the VLA Imaging Pipeline\footnote{\url{https://science.nrao.edu/facilities/vla/data-processing/pipeline/vipl}}. We detected a $\sim 40 \sigma$ source coincident with the optical position of the afterglow of GRB\,250419A \citep[]{2025GCN.40212....1S}. We use \texttt{pwkit/imtool} \citep[]{2017ascl.soft04001W} to measure the flux density of the source, and find $F_\nu =296 \pm 8~\mu$Jy. We initiated further observations at $S$- (central frequency of 3~GHz, bandwidth of 2~GHz), $C$- (central frequency of 6~GHz, bandwidth of 4~GHz), $X$-, $K_{\rm u}$- (central frequency of 15~GHz, bandwidth of 6~GHz), $K$- (central frequency of 22~GHz, bandwidth of 8~GHz), and $K_{\rm a}$-band (central frequency of 33~GHz, bandwidth of 8~GHz), spanning $\delta t \approx 4.1$--$181.8~$days. These observations are summarized in Table~\ref{tab:radioobs}. 

\subsubsection{NOEMA observations}

We observed the position of GRB\,250419A with the NOrthern Extended Millimeter Array (NOEMA) at a mid-time of $\delta t = 10.92,20.83~$days at mean frequencies of $86~$ and $102~$GHz (15.5~GHz bandwidth) under Director's Discretionary Time program E24AN (PI Ho). The first observation was obtained in 12C configuration and the second observation was obtained in the configuration ``9ant'' (intermediate between C and D). In both epochs MWC349 was used as the primary flux calibrator and 3C273 was used as the bandpass calibrator. The data reduction was done with Continuum and Line Interferometer Calibration (\texttt{CLIC}) and \texttt{MAPPING}, which are part of the \texttt{GILDAS} software package \citep{gildas}\footnote{\href{https://www.iram.fr/IRAMFR/GILDAS}{https://www.iram.fr/IRAMFR/GILDAS}}. We detect the millimeter afterglow of GRB\,250419A in both epochs and at both frequencies. These observations are summarized in Table~\ref{tab:radioobs}.

\startlongtable
\begin{deluxetable*}{cccc}
\tablecolumns{4}
\tablecaption{Radio observations of GRB\,250419A 
\label{tab:radioobs}}
\tablehead{\colhead{Observatory} & \colhead{$\delta t^{a}$} & \colhead{$\nu^b$}  & \colhead{Flux Density$^{c}$}
\\ 
\colhead{} & \colhead{(d)} & \colhead{(GHz)} & \colhead{($\mu$Jy)}
}
\startdata
VLA&3.19&10&$296 \pm 8$\\
&4.06&3&$\lesssim 38$\\
&&6&$230 \pm 21$\\
&&10&$363 \pm 15$\\
&&15&$648 \pm 15$\\
&10.18&3&$164 \pm 34$\\
&&6&$616 \pm 13$\\
&&10&$755 \pm 14$\\
&&15&$708 \pm 13$\\
&&22&$663 \pm 18$\\
&&33&$566 \pm 16$\\
&20.29&3&$119 \pm 39$\\
&&6&$147 \pm 17$\\
&&10&$194 \pm 14$\\
&&15&$266 \pm 10$\\
&&22&$247 \pm 28$\\
&&33&$199 \pm 39$\\
&49.12&3&$124 \pm 17$\\
&&6&$195 \pm 12$\\
&&10&$158 \pm 13$\\
&&15&$108 \pm 11$\\
&&22&$69 \pm 14$\\
&&33&$\lesssim 31$\\
&89.0&3&$56 \pm 17$\\
&&6&$91 \pm 14$\\
&&10&$58 \pm 15$\\
&&15&$69 \pm 12$\\
&&22&$\lesssim 51$\\
&181.81&3&$44 \pm 13$\\
&&6&$31 \pm 12$\\
&&10&$\lesssim 19$\\
\hline
NOEMA&10.92&86&$308 \pm 19$\\
&&102&$281 \pm 21$\\
&20.83&86&$159 \pm 26$\\
&&102&$151 \pm 29$\\
\hline
\enddata
\tablenotetext{a}{Midtime of observation compared to SVOM 
trigger}
\tablenotetext{b}{Mid Frequency}
\tablenotetext{c}{Errors quoted are $1 \sigma$, upper limits are $3 \sigma$.}
\end{deluxetable*}

\section{Afterglow Properties and Comparison}
\label{sec:Analysis}

Here we place the afterglow of GRB\,250419A into the context of synchrotron radiation in order to determine basic physical properties of the event. We also compare the prompt emission and afterglow of GRB\,250419A to the broader population of long GRBs and XRFs.

\subsection{Basic Considerations}

We interpret the X-ray, optical, and radio emission following GRB\,250419A in the context of synchrotron emission produced by a relativistic forward shock originating from the collimated jet interacting with the surrounding environment 
\citep{1998ApJ...497L..17S, 1999ApJ...523..177W, 2000ApJ...543...66P, 2002ApJ...568..820G}. In this scenario, the electrons are accelerated into a nonthermal powerlaw distribution ($N(\gamma_e) \propto \gamma_e^{-p}$), and the emission can be described by the peak flux density ($F_{\nu, \rm p}$) and three break frequencies: the self absorption frequency ($\nu_{\rm sa}$), the characteristic frequency ($\nu_{\rm m}$), and the cooling frequency ($\nu_{\rm c}$). The behavior of the afterglow is dependent on the  isotropic kinetic energy of the blast wave ($E_{\rm K, iso}$), the circumburst density of the GRB, and the fraction of energy imparted on the electrons ($\varepsilon_{\rm e}$) and the magnetic field ($\varepsilon_{\rm B}$). The circumburst density is either assumed to be constant with radius (ISM-like, $\rho = n_0 m_p$) or follows a wind-like profile ($\rho \propto r^{-2}$). As the GRB jet decelerates, the beaming angle ($\theta_{\rm beam} = 1/\Gamma$, where $\Gamma$ is the bulk Lorentz factor) will approach the size of the opening angle of the jet ($\theta_{\rm j}$). At this time ($t_{\rm jet}$), an achromatic break in the afterglow (known as the ``jet break'') is expected, and the emission will follow $F_\nu \propto t^{-p}$ for $\nu_{\rm m} < \nu_{\rm obs}$, where $\nu_{\rm obs}$ is the observing frequency \citep{1999ApJ...525..737R, 1999ApJ...519L..17S}.

In what follows, we assume a constant density ISM-like environment, as many long GRBs are well fit with ISM-like environments \citep[e.g.][]{2021ApJ...911...14K, 2022ApJ...940...53S, 2026arXiv260328699L}. Additionally, it is expected that the afterglow behavior after $t_{\rm jet}$ is agnostic to the density profile of the surrounding environment \citep{1999ApJ...519L..17S}, and the majority of our data was taken $\delta t > t_{\rm jet}$ (Section~\ref{sec:xray_optical}).  We also assume the standard slow-cooling break frequency ordering of $\nu_{\rm sa} < \nu_{\rm m} < \nu_{\rm c}$, though we note the early afterglow may be in the fast-cooling regime.

\begin{figure*}
    \centering
    \includegraphics[width = 0.97\textwidth]{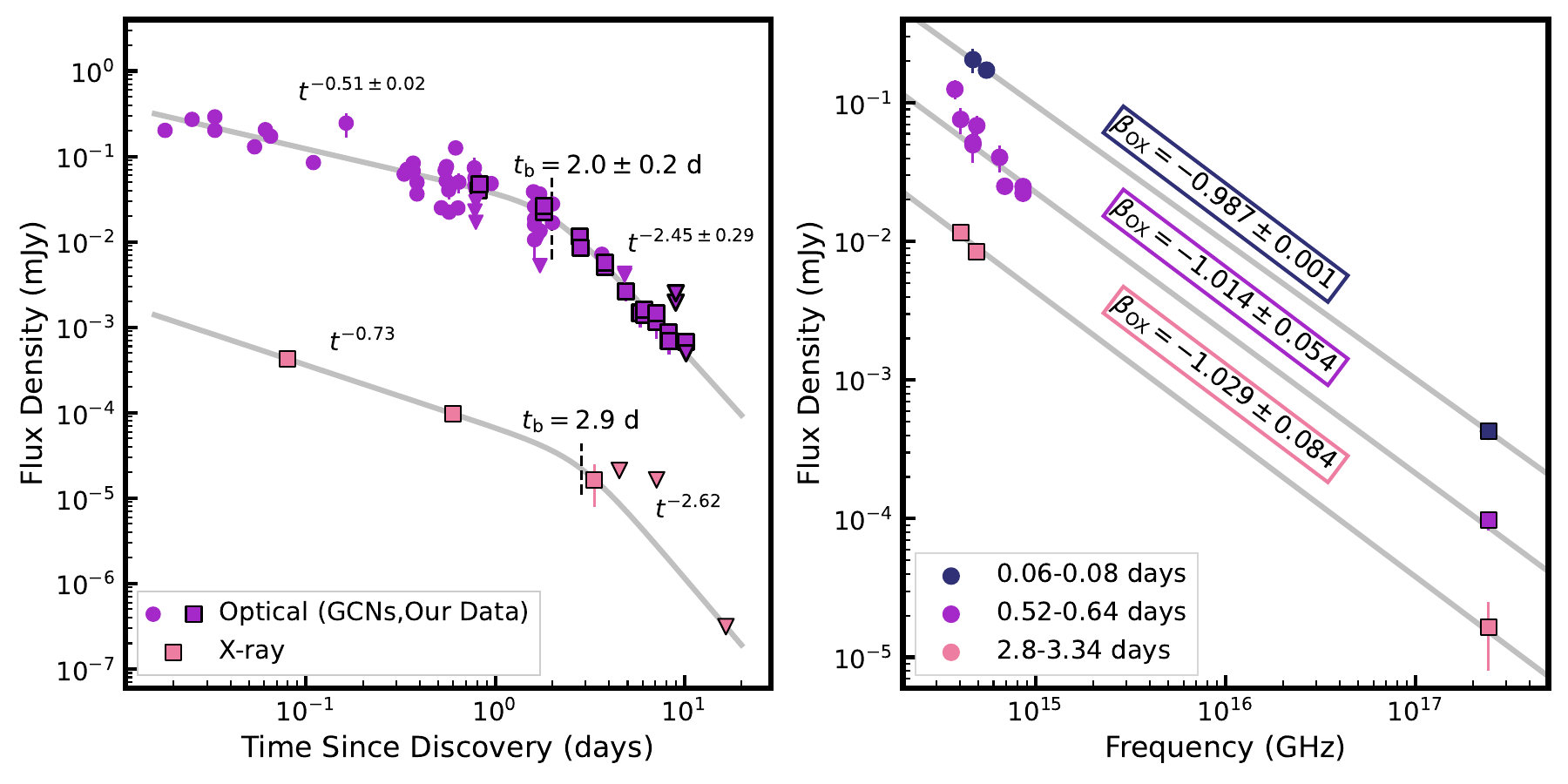}
    \caption{{\it Left:} The optical (purple) and X-ray (pink) light curves of the afterglow of GRB\,250419A. Squares represent data newly presented here, whereas circles represent data from the General Coordinates Network (GCN). The data are fit to a smoothly broken power law (grey), with the pre- and post-break decline indices and break time ($t_{\rm b}$) indicated. Triangles represent non-detections. {\it Right:} The optical and X-ray spectral energy distributions at three common epoches. The data are fit to a power-law, and the optical to X-ray spectral index ($\beta_{\rm OX}$) is indicated for each epoch.}
    \label{fig:XrayOptical}
\end{figure*}

\begin{figure*}
    \centering
    \includegraphics[width = 0.97\textwidth]{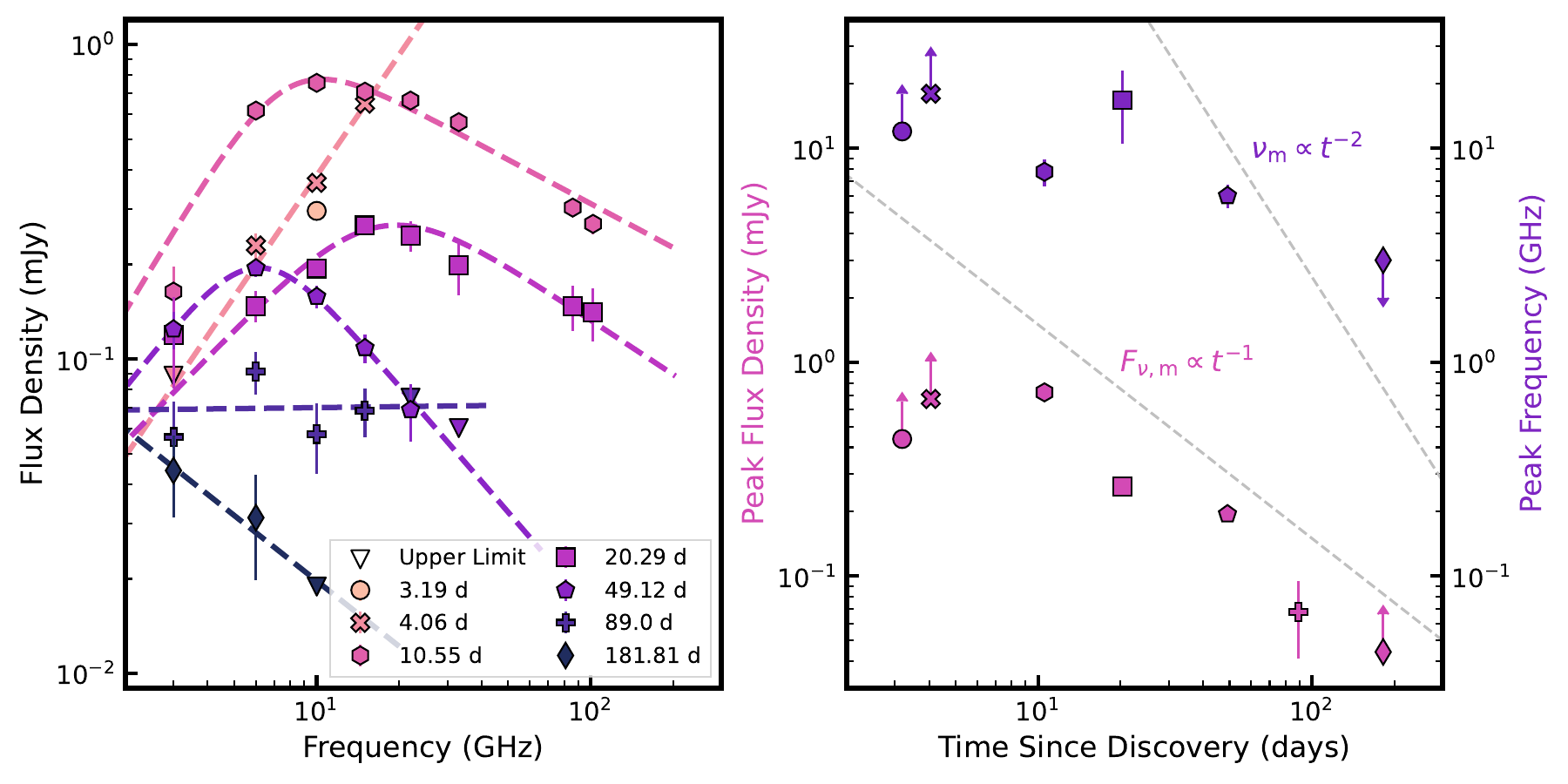}
    \caption{{\it Left:} The radio spectral energy distributions (SEDs) of the afterglow of GRB\,250419A across seven epochs ($\delta t \approx 3.2$--$181.8~$d). For six of the epochs, a single power law or smoothly broken power law is fit to the data, with the spectral index/indices, peak frequency ($\nu_{\rm p}$), and peak flux density ($F_{\nu, \rm p}$) presented in Table~\ref{tab:radiosed}. Triangles represent non-detections. {\it Right}: $F_{\nu, \rm p}$ (pink) and $\nu_{\rm p}$ (purple) of the radio SEDs vs. time. Arrows represent lower or upper limits on the fitted values. To guide the eye, the expected evolution of $F_{\nu, \rm m}$ and $\nu_{\rm m}$ post-jet break \citep{1999ApJ...519L..17S} are also plotted (grey).}
    \label{fig:RadioSED}
\end{figure*}

\subsubsection{Optical and X-ray Afterglow}
\label{sec:xray_optical}

We first consider the optical and X-ray afterglow in order to constrain $t_{\rm jet}$, $\theta_{\rm j}$, and the location of $\nu_{\rm m}$, $\nu_{\rm c}$.
We find that the optical afterglow is well described by a smoothly broken powerlaw with $\alpha_1 = -0.52 \pm 0.02$ and $\alpha_2 = -2.4 \pm 0.3$, and a break time of $t_{\rm b} = 2.0 \pm 0.2~$days (Figure~\ref{fig:XrayOptical}). Similarly, thanks to the deep {\it Chandra} non-detection at $\delta t \approx 16.5~$days, the X-ray afterglow can be described as a broken powerlaw with $\alpha_1 \approx -0.7$ and $\alpha_2 \lesssim -2.6$, and a break time of $t_{\rm b} \approx 2.9~$days, though we note due to the scarcity of X-ray data there are many degeneracies in the fit. Given the similarity between the optical and X-ray post-break temporal indices, as well as the agreement with the break time, we consider this break to be the jet break, and estimate $p \approx 2.4$--$2.6$ and $t_{\rm jet} \approx 2.0$--$2.9~$day, and also determine $\nu_{\rm m} < \nu_{\rm O}, \nu_{\rm X}$ at $\delta t \gtrsim 3~$days. We can use this measurement of $t_{\rm jet}$ to determine $\theta_{\rm j}$, following \citet{1999ApJ...519L..17S}:
\begin{equation}
    \frac{t_{\rm jet}}{\rm day} \approx 0.26(1+z)\left(\frac{E_{\rm K, iso}}{10^{52}~{\rm erg}} \frac{1~{\rm cm}^{-3}}{n_0}\right)^{1/3}\left(\frac{\theta_{\rm j}}{0.1~{\rm rad}}\right)^{8/3}\\
\end{equation}
The $E_{\rm K, iso}$ and $n_0$ term has a weak dependence on the opening angle, and we estimate it to be unity. As a result, $t_{\rm jet} \approx 2.0$--$2.9~$day leads to $\theta_{\rm j} \approx 0.17$--$0.20~$rad ($9.7$--$11.5^{\circ}$).

We next use the estimate of $p$ to determine the position of $\nu_{\rm c}$. For $p \approx 2.5$, the optical to X-ray spectral index is expected to be $\beta_{\rm OX} = -p/2 \approx -1.25$ if $\nu_{\rm c} < \nu_{\rm O} < \nu_{\rm X}$ and $\beta_{\rm obs} = (1-p)/2 \approx -0.75$ if $\nu_{\rm O}, \nu_{\rm X} < \nu_{\rm c}$, whereas if $\nu_{\rm O} < \nu_{\rm c} <  \nu_{\rm X}$ then $\beta_{\rm OX}$ should be somewhere between $-0.75$ and $-1.25$. Between $\sim 0.07~$--$3~$days, we measure $\beta_{\rm OX} \approx -1$ (Figure~\ref{fig:XrayOptical}), indicating that $\nu_{\rm O} < \nu_{\rm c} <  \nu_{\rm X}$. In this regime, the expected X-ray decline pre-jet break is $\approx -1.4$ and the expected optical decline is $\approx -1.1$ for an ISM-like density profile\footnote{The expectation for a wind-like environment is an optical decline of $\approx -1.6$, even steeper than the ISM-like density profile.}. The measured pre-break declines of both the X-ray and optical afterglows ($\approx -0.7$ and $\approx -0.5$, respectively) are much shallower than these expectations. We return to this point in Section~\ref{sec:opt_comparison}. 

\subsubsection{Radio Afterglow}
\label{sec:radio_afterglow}

We next explore the radio spectral energy distributions (SEDs) in order to determine whether $\nu_{\rm m}$ and $F_{\nu, \rm m}$ can be constrained by the radio data, and whether the radio SEDs follow the expected forward shock evolution. All of the radio data was taken after our estimated $t_{\rm jet}$, and as such it is expected that $\nu_{\rm m} \propto t^{-2}$ and $F_{\nu, \rm m} \propto t^{-1}$. The SEDs at $\delta t = 3.19, 4.06~$days can each be fit with a single power law with a positive spectral index\footnote{see Section~\ref{sec:scintillation} for the $\sim 3.19~$day power law fit}, providing lower limits on the peak flux density and frequency ($F_{\nu, \rm peak}$ and $\nu_{\rm peak}$). The SEDs at $\delta t$ = 10.55, 20.29, and 49.12~days can be fit with smoothly broken power laws, constraining the location of $F_{\nu, \rm peak}$ and $\nu_{\rm peak}$. The SED at $\delta t = 89~$days is consistent with a flat spectrum, and therefore no constraint can be placed on $\nu_{\rm peak}$, however we estimate $F_{\nu, \rm peak} \approx 68 \pm 27~\mu$Jy for this epoch. The SED at $\delta t = 181.81~$days can be fit with a single power law with a negative spectral index, providing a lower limit on $F_{\nu, \rm peak}$ and an upper limit on $\nu_{\rm peak}$. The SED fits and evolution of $F_{\nu, \rm peak}$ and $\nu_{\rm peak}$ are displayed in Figure~\ref{fig:RadioSED}, and the summary of the fit is presented in Table~\ref{tab:radiosed}.

\startlongtable
\begin{deluxetable*}{ccccc}
\tablecolumns{5}
\tablecaption{Fits to the radio spectral energy distributions of GRB\,250419A 
\label{tab:radiosed}}
\tablehead{\colhead{$\delta t$} & \colhead{Fit Type$^{a}$} & \colhead{Spectral Index/Indices} & \colhead{$\nu_{\rm peak}$}  & \colhead{$F_{\nu, \rm peak}$}
\\ 
\colhead{(d)} & \colhead{} & \colhead{} & \colhead{(GHz)} & \colhead{($\mu$Jy)} 
}
\startdata
3.19 & PL & $\beta = 2.0 \pm 0.1$ & $\gtrsim 12$ & $\gtrsim 440$ \\
4.06 & PL &  $\beta = 1.3 \pm 0.1$  & $\gtrsim 18$ & $\gtrsim 670$ \\
10.55 & BPL & $\beta_1 = 1.5 \pm 0.5, \beta_2 = -0.5 \pm 0.1$ & $7.8 \pm 1.1$ & $723 \pm 37$ \\
20.29 & BPL &  $\beta_1 = 0.9 \pm 0.3, \beta_2 = -0.6 \pm 0.2$& $16.8 \pm 6.3$ & $263 \pm 23$ \\
49.12 & BPL &  $\beta_1 = 1.1 \pm 0.2, \beta_2 = -1.0 \pm 0.1$ & $6.0 \pm 0.8$ & $195 \pm 5$ \\
89.00 & PL &  $\beta = 0.0 \pm 0.2$  & -- & $\approx 70 \pm 10$ \\
181.81 & PL &  $\beta = -0.7 \pm 0.1$  & $\lesssim 2$ & $\gtrsim 44$ \\
\hline
\enddata
\tablenotetext{a}{Whether each epoch was fit with a power law (PL) or broken power law (BPL)}
\tablenotetext{b}{Errors quoted are $1 \sigma$.}
\end{deluxetable*}

Overall, $F_{\nu, \rm peak}$ approximately follows the expected evolution of $F_{\nu, \rm m}$, however, $\nu_{\rm peak}$ does not follow the expected evolution of $\nu_{\rm m}$. In particular,  $\nu_{\rm peak}$ appears to be much lower than expected at $\delta t \approx 10.55~$days (Figure~\ref{fig:RadioSED}). The deviation from the expected evolution of $\nu_{\rm m}$ indicates that $\nu_{\rm peak}$ is not $\nu_{\rm m}$, and that the radio afterglow is dominated by an additional feature outside of the standard forward shock model at $\delta t \lesssim 10.55~$days. This additional feature does not appear to be scintillation (see Section~\ref{sec:scintillation}), which can cause deviations from the standard forward shock model, and instead is likely the result of an additional emission mechanism, such as a reverse shock \citep{2000ApJ...535L..33S, 2000ApJ...532..286K, 2003ApJ...595..950Z, 2005ApJ...625..263G, 2005ApJ...628..315Z}. At $\delta t \gtrsim 10.55~$days the radio SEDs appears to follow the expected evolution of $\nu_{\rm m}$ and $F_{\nu, \rm m}$.

\subsection{Comparison to Long GRBs and XRFs}

In this section we investigate the prompt emission and afterglow properties of GRB\,250419A to determine whether its properties are consistent with the overall population of traditional on-axis long GRBs, as well as the known population of low-luminosity GRBs (LLGRBs) and XRFs.

\subsubsection{Prompt Emission}

\begin{figure*}
    \centering
    \includegraphics[width = 0.97\textwidth]{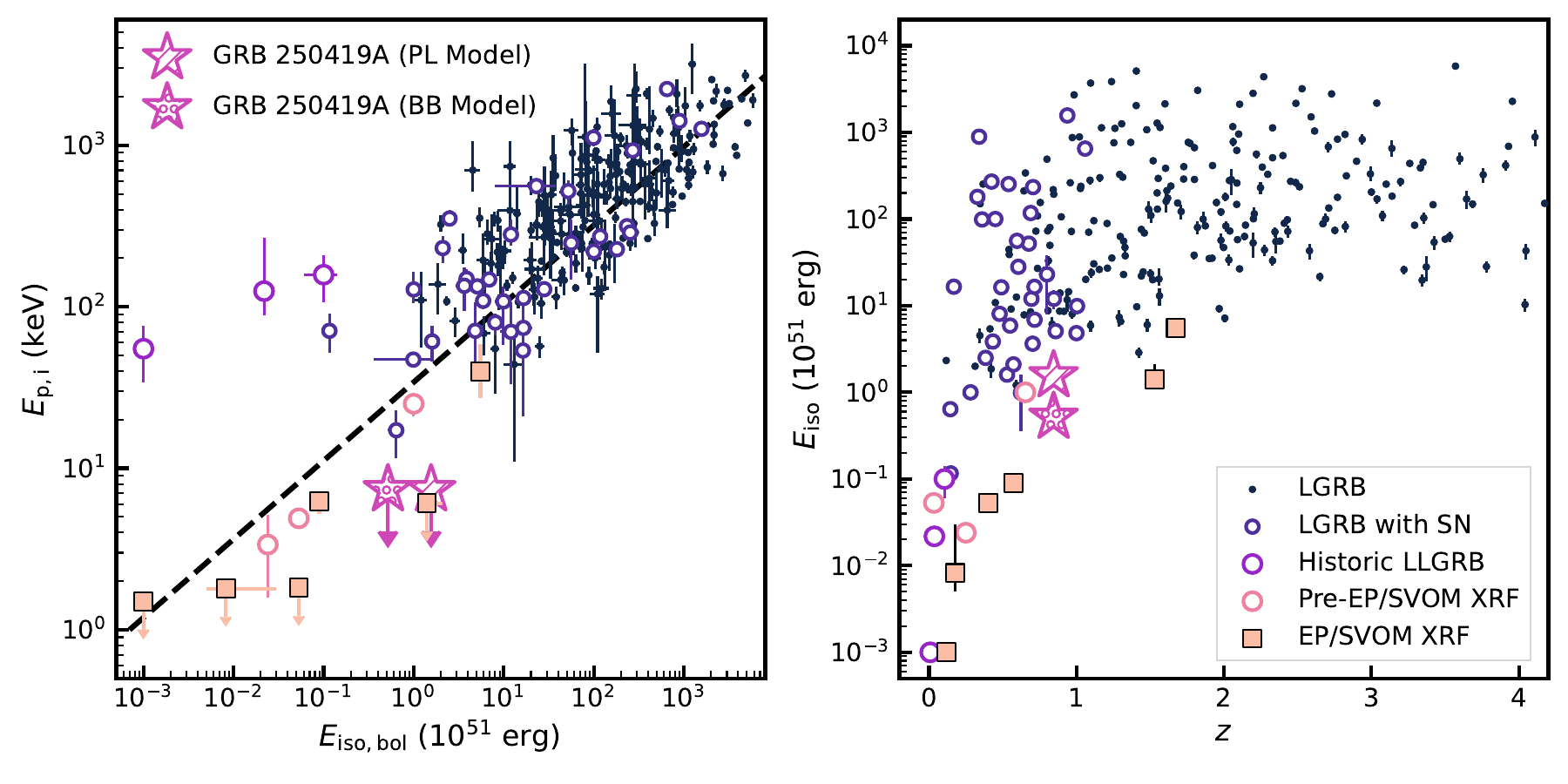}
    \caption{{\it Left:} Redshift corrected peak energy of the prompt emission ($E_{\rm p, i}$) vs bolometric corrected isotropic energy of the prompt emission ($E_{\rm iso, bol}$)  also known as the Amati diagram \citep{2002A&A...390...81A, 2008MNRAS.391..577A}. Here we distinguish between long GRBs with and without detected supernovae (solid and open circles, respectively), pre-EP/SVOM X-ray flashes (XRFs, pink), low-luminosity gamma-ray bursts (LLGRBs, purple), and newly detected XRFs from EP and SVOM (peach squares), with data from \citealt{2020MNRAS.492.1919M, 2025NatAs...9.1073S, 2025arXiv250417034L, 2025ApJ...988L..34J, 2025arXiv251210239S, 2026arXiv260420346S, 2026arXiv260326213D}. To guide the eye, we also plot the ``Amati relation'' which correlates $E_{\rm p, i}$ and  $E_{\rm iso, bol}$. GRB\,250419A is represented as stars for $E_{\rm iso, bol}$ calculated assuming a power-law model (PL, diagonal hash), and a black-body model (BB, circle hash). See Section~\ref{sec:svom_detection} for more details. {\it Right:} $E_{\rm iso, bol}$ vs redshift for the same sample as plotted on the left. }
    \label{fig:Amati}
\end{figure*}

We first investigate the prompt emission of GRB\,250419A. From the SVOM detection (Section~\ref{sec:svom_detection}), the rest-frame peak energy ($E_{\rm p,i} = E_{\rm p} (1+z)$) is $E_{\rm p,i} < 7.4~$keV at a redshift of $z = 0.845$ (Section~\ref{sec:optical_followup}). We calculate the bolometric isotropic $\gamma$-ray energy using the $k_{\rm bol}$ calculated in Section~\ref{sec:svom_detection}, and find for the power-law model $E_{\rm iso, bol, pl} \approx 1.6 \times 10^{51}~$erg ($E_{\rm iso,pl} \approx 2.8 \times 10^{50}~$erg) and for the black-body model we find $E_{\rm iso, bol,bb} \approx 5.2 \times 10^{50}~$erg ($E_{\rm iso,bb} \approx 2.7 \times 10^{50}~$erg). 

We place GRB\,250419A in the context of the larger GRB population, by plotting it in the $E_{\rm p,i}$-$E_{\rm iso, bol}$ phase space, known as the Amati diagram (\citealt{2002A&A...390...81A, 2008MNRAS.391..577A}, see also \citealt{2004ApJ...609..935Y}), as well as the $E_{\rm iso, bol}$-$z$ phase space (Figure~\ref{fig:Amati}). 
While GRB\,250419A appears to be an outlier within the overall population of long GRBs in the Amati diagram (especially if we assume $E_{\rm iso, bol, pl}$), it is difficult to definitively claim this given the under-sampling of the region it resides in. It is also unclear if the relation between $E_{\rm p,i }$ and $E_{\rm iso, bol}$ holds true for $E_{\rm iso, bol} < 10^{52}~$erg.
An alternative parameter space that can be considered is $E_{\rm iso, bol}$ vs. $z$. Regardless of if we assume $E_{\rm iso, bol, pl}$ or $E_{\rm iso, bol, bb}$ for GRB\,250419A, it appears to be an outlier compared to long GRBs at a similar redshift (Figure~\ref{fig:Amati}).
As SVOM and EP continue to detect more events, the under-sampled portions of these parameter spaces should begin to be filled in, and therefore we will better be able to identify outliers in these regimes.

\subsubsection{X-ray Afterglow}
\label{sec:xray_comparison}

\begin{figure*}
    \centering
    \includegraphics[width = 0.9\textwidth]{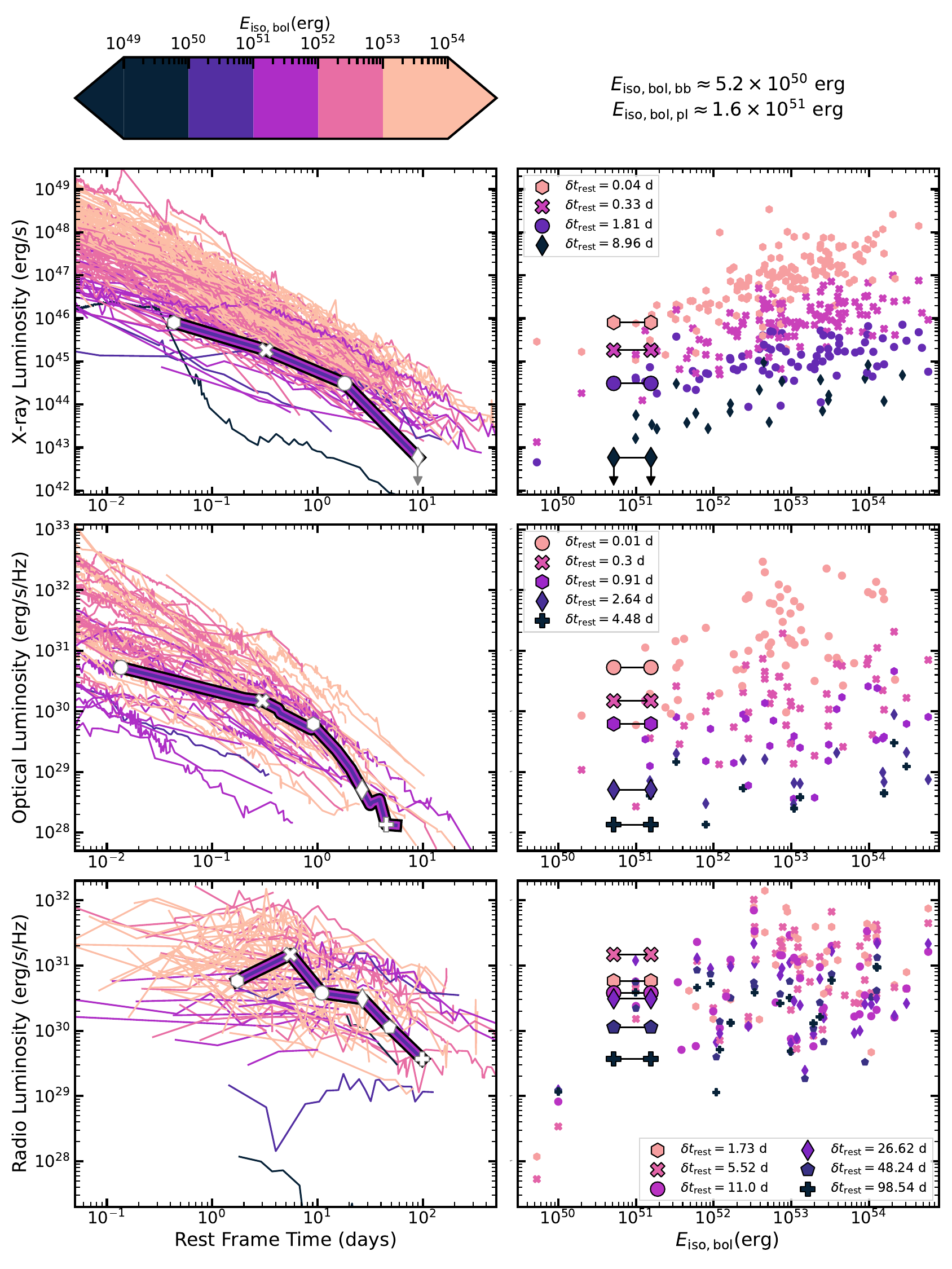}
    \caption{{\it Left:} The X-ray (top), optical (middle), and radio (bottom) afterglows of a sample of long GRBs, colored by the bolometric corrected isotropic energy \citep[$E_{\rm iso, bol}$, ][]{2006ApJ...637..889Z, 2010ApJ...720.1513K} of the prompt emission. GRB\,250419A is highlighted as a thick line with both the power law (pl) and black body (bb) model values of $E_{\rm iso, bol}$. {\it Right:} The luminosity vs. $E_{\rm iso, bol}$ phase space for specific data points of GRB\,250419A, compared to the overall long GRB population at similar rest frame times. GRB\,250419A is similar in luminosity compared to most long GRBs, however the emission may be slightly underluminous for the measured $E_{\rm iso, bol}$ at some points in the light curves. }
    \label{fig:Lum_vs_Eiso}
\end{figure*}

We next compare the X-ray afterglow of GRB\,250419A to the overall long GRB population. We display the X-ray light curves of the long GRB population in Figure~\ref{fig:Lum_vs_Eiso} (top left), and color the light curves by the $E_{\rm iso,bol}$ \citep{2010ApJ...720.1513K, 2020MNRAS.492.1919M}. We additionally sample the long GRB X-ray light curves at similar rest-frame times of our observations of GRB\,250419A ($\delta t_{\rm rest} \approx 0.04, 0.33, 1.81, 8.96~$days). In the cases where there is not a reasonable X-ray detection within $\sim 10\%$ of the stated epochs, we interpolate the X-ray light curve by fitting a power law to the nearest 4 data points. We then plot the X-ray luminosity at each epoch against the $E_{\rm iso, bol}$ of the GRB. The results are displayed in Figure~\ref{fig:Lum_vs_Eiso} (top right), and are colored by epoch. 

At $z = 0.845$, the X-ray afterglow of GRB\,250419A is first detected at $\sim 8 \times 10^{45}~{\rm erg~s}^{-1}$ ($\delta t_{\rm rest} \approx 0.04~$d), and fades to $\lesssim 6 \times 10^{42}~{\rm erg~s}^{-1}$ ($\delta t_{\rm rest} \approx 8.96~$d). Ignoring $E_{\rm iso, bol}$, the X-ray afterglow of GRB\,250419A is on the lower end of the luminosity distribution compared to the overall long GRB population. However, assuming $E_{\rm iso, bol, bb} \approx 5.2 \times 10^{50}~$erg the X-ray afterglow is somewhat more luminous than GRBs with similar $E_{\rm iso, bol}$ at $\delta t_{\rm rest} \lesssim 0.33~$d, and less luminous than GRBs with similar $E_{\rm iso, bol}$  for $\delta t_{\rm rest} \gtrsim 8.96~$d. If we instead assume $E_{\rm iso, bol, pl} \approx 1.6\times 10^{51}~$erg, the X-ray afterglow of GRB\,250419A is at a typical luminosity compared to GRBs with a similar $E_{\rm iso, bol}$ for $\delta t_{\rm rest} \lesssim 1.81~$d, but is more starkly under-luminous at $\delta t_{\rm rest} \gtrsim 8.96~$d.

\subsubsection{Optical Afterglow}
\label{sec:opt_comparison}

Similar to the X-ray afterglow, we compare the optical afterglow of GRB\,250419A to the overall long GRB population. We plot the optical light curve of the long GRB population in Figure~\ref{fig:Lum_vs_Eiso} (middle left), and color the light curves by the $E_{\rm iso,bol}$ \citep{2010ApJ...720.1513K, 2020MNRAS.492.1919M}. We additionally sample the long GRB optical light curves at a few select rest-frame times of our observations of GRB\,250419A ($\delta t_{\rm rest} \approx 0.01, 0.30, 0.91, 2.64, 4.48~$days), and handle interpolating the optical afterglows the same way as we did for the X-ray afterglows. The results are displayed in Figure~\ref{fig:Lum_vs_Eiso} (middle right), and are colored by epoch.

At $z =0.845$, the optical afterglow of GRB\,250419A is first detected at $\sim 5 \times 10^{30}~{\rm erg~s}^{-1}~{\rm Hz}^{-1}$ ($\delta t_{\rm rest} \approx 0.01~$d) and fades to $\sim 10^{28}~{\rm erg~s}^{-1}~{\rm Hz}^{-1}$ ($\delta t_{\rm rest} \approx 5.5~$d). Ignoring $E_{\rm iso, bol}$, the optical afterglow of GRB\,250419A is well within the luminosity distribution of the long GRB population. However, assuming $E_{\rm iso, bol, bb}$ the optical afterglow is somewhat more luminous than GRBs with similar $E_{\rm iso, bol}$ at $\delta t_{\rm rest} \lesssim 1~$d. Indeed, using Equation 1 from
\citet{2010ApJ...720.1513K}, we would expect the optical afterglow of GRB\,250419A to be $\sim 3 \times$ less luminous at $\delta t_{\rm rest} = 1~$d. At $\delta t_{\rm rest} \gtrsim 1~$d the optical afterglow luminosity is more consistent with the overall long GRB population, and may even be underluminous at $\delta t_{\rm rest} \gtrsim 3~$d. If we instead assume $E_{\rm iso, bol, pl}$, the optical afterglow of GRB\,250419A is at a typical luminosity compared to GRBs with a similar $E_{\rm iso, bol}$ for $\delta t_{\rm rest} \lesssim 3~$d, and is somewhat underluminous at $\delta t_{\rm rest} \gtrsim 3~$d.

\begin{figure*}
    \centering
    \includegraphics[width = 0.97\textwidth]{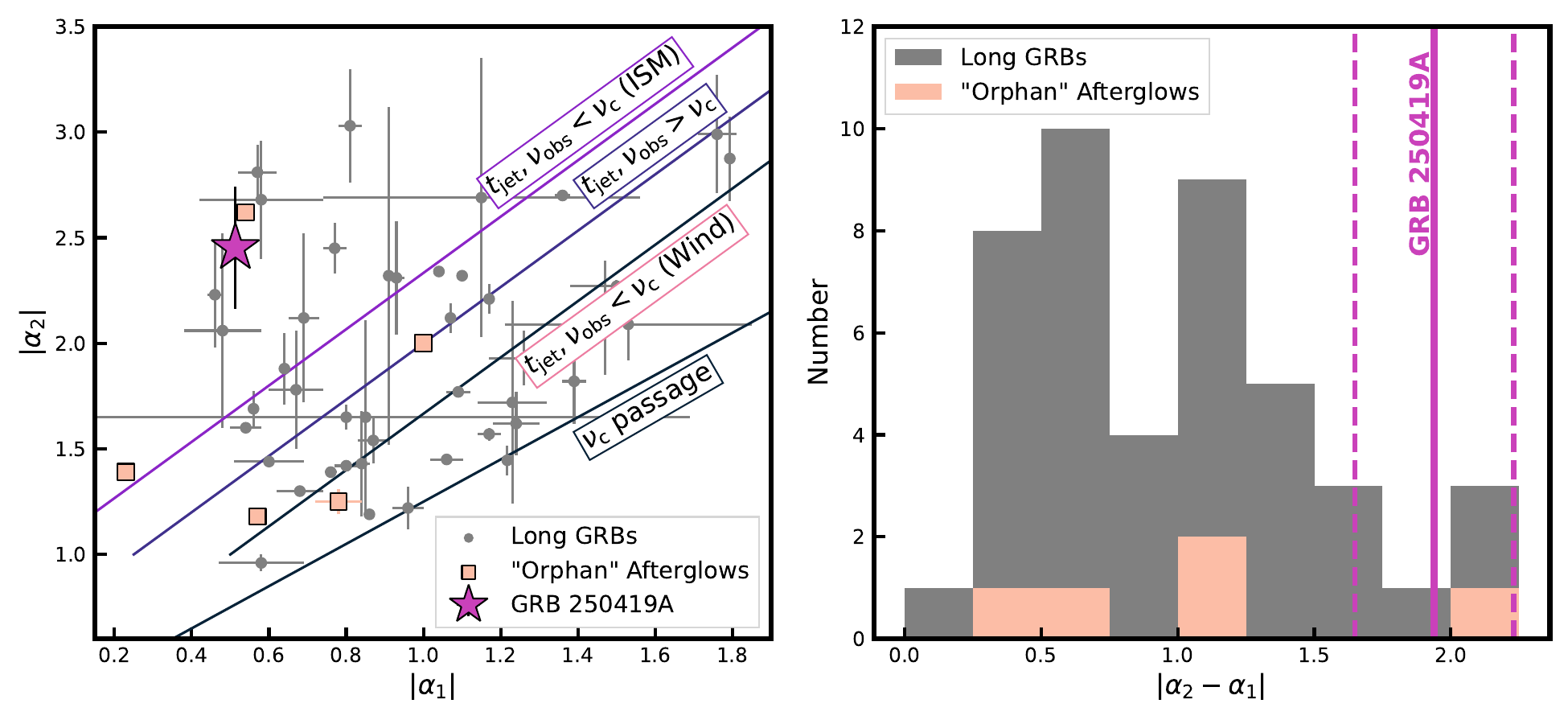}
    \caption{{\it Left:} Optical post-break temporal index ($|\alpha_2|$) vs pre-break temporal index ($|\alpha_1|$) for long GRBs \citep[grey,][]{2006ApJ...637..889Z, 2010ApJ...720.1513K}, ``orphan afterglows'' \citep[peach squares,][]{2021ApJ...918...63A, 2020ApJ...905...98H, 2022ApJ...938...85H, 2021ApJ...918...63A, 2025ApJ...985..124L, 2025MNRAS.537.2362P}, and GRB\,250419A (pink star). Also shown are the expectations for $|\alpha_2|$ vs $|\alpha_1|$ for the passage of $\nu_{\rm c}$, and the onset of a jet break (lines, \citealt{2002ApJ...568..820G, 1999ApJ...519L..17S}), for various break frequency orderings, $p$, and explosion environments (ISM vs wind). {\it Right:} $|\alpha_2-\alpha_1|$ for the same sample shown on the left.}
    \label{fig:alpha1_alpha2}
\end{figure*}

In addition to the optical luminosity, we can compare the pre- and post break slopes of the optical afterglow to the overall population. As mentioned in Section~\ref{sec:xray_optical}, the optical afterglow can be fit with a broken power-law with temporal indices of $\alpha_1 = -0.51 \pm 0.02$ and $\alpha_1 = -2.45 \pm 0.29$, resulting in $|\Delta \alpha| = |\alpha_2-\alpha_1| = 1.94 \pm 0.29$. 

We compare the values of $\alpha_1$, $\alpha_2$, and $\Delta \alpha$ to the overall long GRB population in Figure~\ref{fig:alpha1_alpha2} \citep{2006ApJ...637..889Z, 2010ApJ...720.1513K}. We additionally include five optically discovered afterglow candidates with clear breaks in the optical light curves \citep[AT2019pim, AT2020blt, AT2021any, AT2021buv, and AT2023lcr,][]{2021ApJ...918...63A, 2020ApJ...905...98H, 2022ApJ...938...85H, 2021ApJ...918...63A, 2025ApJ...985..124L, 2025MNRAS.537.2362P, 2022MNRAS.513.2777K, 2024ApJ...972..162J}, four of which are considered ``orphan'' afterglows\footnote{AT2021buv is considered to be the optical afterglow of GRB\,210204A \citep{2021ApJ...918...63A, 2022ApJ...938...85H, 2022MNRAS.513.2777K, 2024ApJ...972..162J}} as no GRB was detected coincident with the optical transient. We note that AT2021buv can actually be fit with a double broken power-law \citep{2024ApJ...972..162J}, and therefore use $\alpha_2$ and $\alpha_3$ for our subsequent analysis.

To determine whether any events are outliers in the $\alpha_2$-$\alpha_1$ phase space, we plot lines of varying $p$ for the following light curve transitions: either the passage of $\nu_{\rm c}$ for an ISM or wind environment ($\alpha_1 = 3(1-p)/4 \rightarrow \alpha_2 = (2-3p)/4$ or $\alpha_1 = (2-3p)/4 \rightarrow \alpha_2 = (1-3p)/4$, respectively), or the onset of a jet break ($\alpha_1 = 3(1-p)/4, (1-3p)/4$, or $(2-3p)/4$ and  $\alpha_2 = -p$) in Figure~\ref{fig:alpha1_alpha2} (left). Compared to the overall long GRB population, GRB\,250419A is an outlier in the $\alpha_2$-$\alpha_1$ parameter space, and its light curve is inconsistent with the expectations of a standard forward shock afterglow (Section~\ref{sec:xray_optical}). Other notable outliers are: GRBs\,990510, 000301, 030226, 030429, 050319, and 050408. The inconsistency within the $\alpha_2$-$\alpha_1$ parameter space can easily be explained for GRB\,000301 \citep[where the shallow $\alpha_1 \approx 0.57$ is the result of a re-brightening event likely due to microlensing][]{2000ApJ...544L..11G}, and GRBs\,030226, 030429, 050319 \citep[which all demonstrated a re-brightening events/plateaus][]{2003ApJ...591L..21D,  2004A&A...427..785J, 2006ApJ...637..889Z,  2010ApJ...720.1513K}. The remaining two GRBs are GRB\,990510, for which no explanation is given to reconcile the measured $\alpha_1 = 0.77 \pm 0.03$ and $\alpha_2 = 2.45 \pm 0.12$ \citep{2010ApJ...720.1513K}, and GRB\,050408, with $\alpha_1 = 0.48 \pm 0.10$ and $\alpha_2 = 2.06 \pm 0.46$. GRB\,050408 was considered an ``X-ray rich'' GRB given its low $E_{\rm p} \approx 20~$keV (observer frame), and one explanation for the shallow $\alpha_1$ is that the GRB was observed moderately off-axis \citep{2007A&A...462L..57D, 2015ApJ...799....3R}. Additionally, of the five optically discovered afterglow candidates, one (AT2020blt) is an outlier, with $\alpha_1 \approx -0.6$ and $\alpha_2 \approx -2.6$, which may be consistent with a slightly off-axis GRB \citep{2020ApJ...905...98H}.

We additionally place GRB\,250419A in the context of the overall long GRB sample by looking at the value of $|\Delta \alpha|$. From standard forward shock expectations, typical values of $|\Delta \alpha|$ for optical afterglows should range from $|\Delta \alpha| \approx 0.25$ (due to the passage of $\nu_{\rm c}$ through the optical band) to $|\Delta \alpha| \approx 1.75$ (due to the onset of a jet break), similar to the distribution seen in Figure~\ref{fig:alpha1_alpha2} (right). Therefore, our measured $|\Delta \alpha| \approx 2$ for GRB\,250419A is on the high end of the distribution, further placing as an outlier compared to other long GRBs. The long GRBs with similarly high $\Delta \alpha > 1.75$ are GRBs\,000301, 030226, 030429, 050319, for which we have explained the optical behavior above. AT2020blt also has a high $\Delta \alpha \approx 2.1$, making it the most similar to GRB\,250419A in both the $\alpha_2$-$\alpha_1$ parameter space and in $\Delta \alpha$. Multiwavelength data of AT2020blt was not sufficiently detailed to determine whether the event was truly off-axis, or just an on-axis GRB that was missed by high-energy satellites \citep{2020ApJ...905...98H, 2022MNRAS.512.1391S, 2025ApJ...985..124L}

\subsubsection{Radio Afterglow}
\label{sec:radio_comparison}

We now compare the radio afterglow of GRB\,250419A to the overall long GRB population. We plot the radio light curves of the long GRB population in Figure ~\ref{fig:Lum_vs_Eiso} (bottom left), and color the light curves by $E_{\rm iso,bol}$ \citep{2010ApJ...720.1513K, 2020MNRAS.492.1919M}. We additionally sample the long GRB radio light curves at the rest-frame times of our observations of GRB\,250419A ($\delta t_{\rm rest} \approx 1.73, 5.52, 11.00, 26.62, 48.24, 98.54~$d\footnote{We omit $\delta t_{\rm rest} \approx 2.20~$d as it is similar to $\delta t_{\rm rest}\approx 1.73~$d.}), and handle interpolating the radio afterglows the same way as explained for the X-ray afterglows (Section~\ref{sec:xray_comparison}). The results are displayed in Figure~\ref{fig:Lum_vs_Eiso} (bottom right) and colored by epoch. 

At $z =0.845$, the radio afterglow of GRB\,250419A is first detected at $\sim 6 \times 10^{30}~{\rm erg~s}^{-1}~{\rm Hz}^{-1}$ ($\delta t_{\rm rest} \approx 1.7~$d), rises to $\sim 1 \times 10^{31}~{\rm erg~s}^{-1}~{\rm Hz}^{-1}$ ($\delta t_{\rm rest} \approx 5.5~$d) and fades to $\sim 4 \times 10^{29}~{\rm erg~s}^{-1}~{\rm Hz}^{-1}$ ($\delta t_{\rm rest} \approx 98.5~$d). Ignoring $E_{\rm iso, bol}$, the radio afterglow of GRB\,250419A is of a similar luminosity to the overall long GRB population. While the X-ray and optical afterglows of long GRBs typically follow a power-law decay (though flares and supernovae may also be observed), radio afterglows often exhibit a rise and decay, making the comparison in Luminosity-$E_{\rm iso, bol}$ space less straightforward. Assuming $E_{\rm iso, bol, bb}$, GRB\,250419A appears to be somewhat over-luminous at $\delta t \lesssim 6~$d, and is otherwise a typical luminosity for a long GRB at a similar $E_{\rm iso, bol}$. Assuming $E_{\rm iso, bol,pl}$, the radio afterglow of GRB\,250419A is at a similar luminosity as other long GRBs with comparable $E_{\rm iso, bol}$ for the majority of the radio afterglow.

\subsection{Radio Constraints on the Lorentz Factor}

\label{sec:scintillation}
\begin{figure*}
    \centering
    \includegraphics[width = 0.98\textwidth]{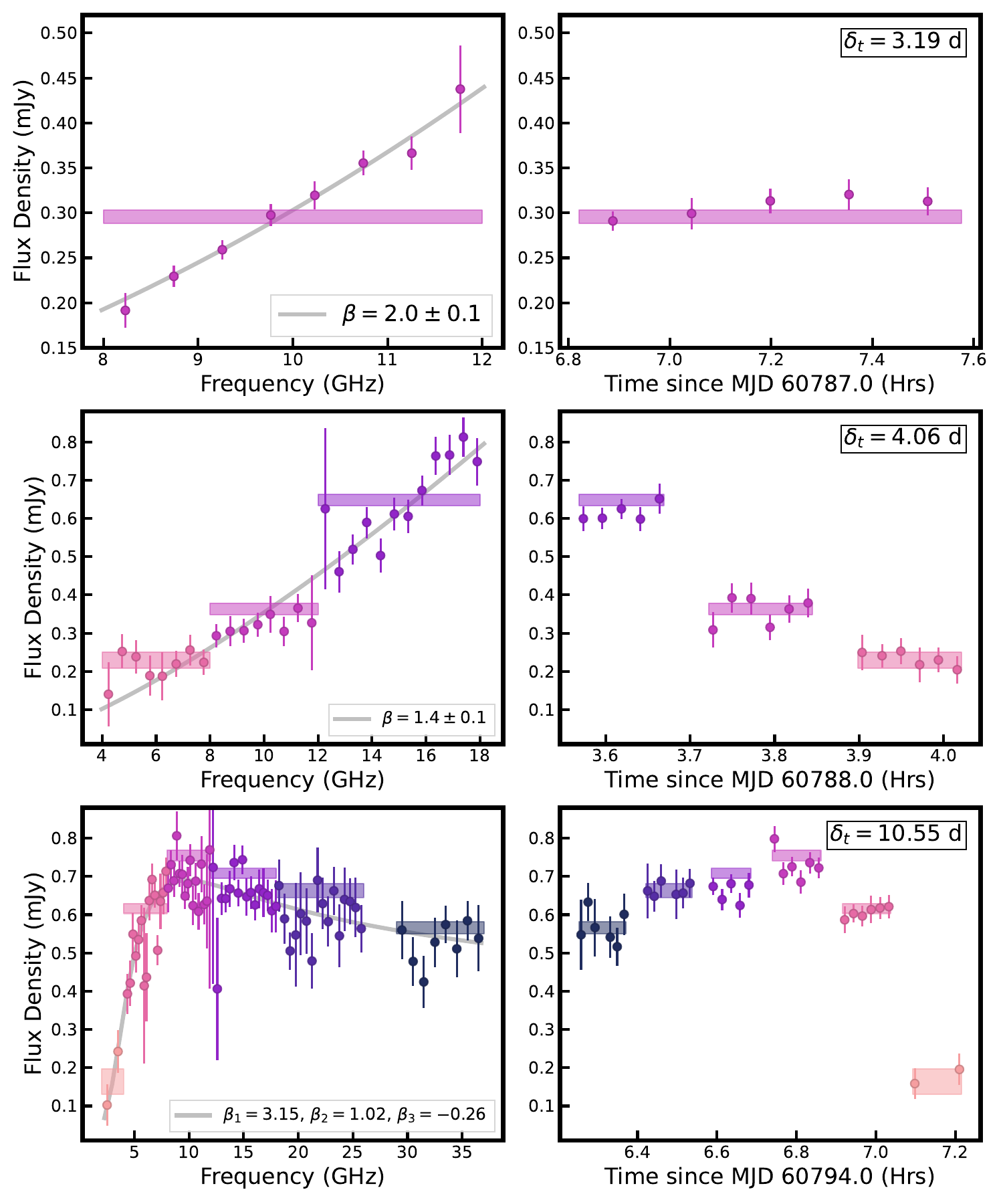}
    \caption{{\it Left:} Radio spectral energy distributions (SEDs) of the afterglow of GRB\,250419A across three epochs ($\delta t \approx 3.06$, top; $\delta t \approx 4.06~$d, middle; and $\delta t \approx 10.55~$d, bottom). The shaded regions are the measured flux density across the entire bandwidth for each radio band. The points represent the flux density for the radio data binned into smaller frequency bins, in order to search for evidence of scintillation. Also shown are power-law and double broken power-law fits to the data (grey) with the resulting spectral indices indicated. {\it Right:} Radio light curves of the afterglow of GRB\,250419A across the same three epochs as shown on the left. Shaded regions are the measured flux density across the entire bandwidth for each radio band. The points represent the flux density for the radio data binned into smaller time bins, in order to search for evidence of scintillation.}
    \label{fig:Scintillation}
\end{figure*}

Here, we utilize the radio detections of GRB\,250419A in order to place constraints on the minimum $\Gamma_0$. From the earliest radio afterglow detection, we can place limits on $\Gamma_0$ using brightness temperature arguments \citep{1998Natur.395..663K, 1999Natur.398..394G}, as has been done for previous GRBs with radio afterglow detections \citep[e.g.][]{2014MNRAS.440.2059A, 2018MNRAS.473.1512A, 2024ApJ...975L..13A, 2025ApJ...994....5A, 2026arXiv260319047C}.
Using Equation 1 from \citet{2026arXiv260319047C}, we use our 10~GHz detection at $\delta t \approx 3.19~$d of $F_{\nu} \approx 0.3~$mJy to calculate $\Gamma_0 \gtrsim 4$. Typically, very early radio observations ($\delta t \lesssim 1~$d) are required to place significant constraints on $\Gamma_0$ utilziing this method \citep{2018MNRAS.473.1512A}. 

Another method for placing a limit on $\Gamma_0$ is through searching for evidence of interstellar scintillation (ISS), which occurs when the emitting region is small enough to be affected by electrons along the line of sight \citep{1990MNRAS.244...68R, 1997NewA....2..449G}. Scintillation can manifest in the large scale modulations of the radio afterglow, observable across frequency and time.
We first search for evidence of scintillation by binning the first three epochs ($\delta t \approx 3.19, 4.06, 10.55~$d) of VLA data into smaller frequency and temporal bins, and measure the flux densities of the binned data, to determine if any large scale variability is present. We present the results of the binning in Figure~\ref{fig:Scintillation}. In the spectrally binned data, there is no obvious signatures of scintillation in the data, as the flux densities of the binned data match the flux density derived from the full bandwidth. Additionally, the binned data is well behaved and can be fit by either a single power-law ($\delta t \approx 3.19, 4.06~$d) or a double broken power law ($\delta t \approx 10.55~$d). Similarly, in the temporally binned data there is no evidence of scintillation, and the flux densities of the binned data match the flux density derived from the full observation window for each band. 

We can utilize the lack of scintillation to place a lower limit on $\Gamma_0$, as it appears that by $\delta t \approx 3.19~$d the emitting region was sufficiently large enough to be unaffected. 
Using NE2025 \citep{2026ApJ..1002....3O}, the critical frequency for ISS at the position of GRB\,250419A is $\approx 7.4$~GHz and an effective screen distance of $D \approx 0.75~$kpc, corresponding to a Fresnel zone size of $\theta_{\rm F0} \approx 3.4 \mu$arcsec, or a physical scale of $8.2\times 10^{16}~$cm \citep{1998MNRAS.294..307W}, assuming an angular distance scale of $\sim 8.5~$kpc/arcsec at $z = 0.845$. At $\delta t \approx 4.06~$d (the first time our observations encompass the transition frequency), this corresponds to $\Gamma_{\rm 4.1~d} \gtrsim 14$. We can extrapolate this limit on $\Gamma$ to an earlier time, first by considering the jet break at $\delta t \approx 2.0$--$2.9~$d (Section~\ref{sec:xray_optical}). We adopt the earliest jet break time of $t_{\rm jet} \approx 2~$d, and assume the angular size of the emitting region is of size $\theta_{\rm GRB, 4.1~d} \approx \theta_{\rm F0}$ at $\delta t \approx 4.06~$d. We assume $\theta_{\rm GRB} \propto t^{1/4}$ \citep{2005ApJ...618..413G} at $\delta t \gtrsim t_{\rm jet}$, and find $\theta_{\rm GRB, 2.0~d} \approx 2.8~\mu$arcsec, corresponding to $\Gamma_{\rm 2.0~d}\gtrsim 24$. Extrapolating $\theta_{\rm GRB, 2.0~d}$ back to the time of the first afterglow detection ($\delta t \approx 0.02~$d), and assuming $\theta_{\rm GRB} \propto t^{5/8}$ (for an ISM-like environment), we find $\theta_{\rm GRB, 0.02~d} \approx 0.1~\mu$arcsec, corresponding to $\Gamma_{\rm 0.02~d}\gtrsim 143$.

In summary, our prompt emission analysis indicates GRB\,250419A is an outlier when compared to the overall population of long GRBs. However, the X-ray, optical, and radio emission is consistent with the overall long GRB population, although it may be slightly under-luminous. The optical emission does demonstrate a shallower than typical pre-break decay, and appears to be inconsistent with expectations from the forward shocks model.
Using the radio afterglow, we place limits on the initial Lorentz factor of $\Gamma_0 \gtrsim 14$, and with some assumptions we can place stronger limits of $\Gamma_0 \gtrsim 143$.

\startlongtable
\begin{deluxetable*}{cccc}
\tablecolumns{4}
\tablecaption{Model free-parameters and prior ranges, conditions and posterior median and 1$\sigma$ confidence intervals.
\label{tab:redback}}
\tablehead{\colhead{Parameter$^{a}$} & \colhead{Prior$^{b}$} & \colhead{Condition} & \colhead{Posterior}
}
\startdata
\multicolumn{4}{c}{Model 1 - Forward Shock Only, On-axis}\\
\hline
$p$ & $2.0 , 2.97$ & -- & $2.61^{+0.05}_{-0.04}$ \\
$\log_{10} n_0$ & $-10,10$ & -- & $-0.63^{+0.09}_{-0.1}$ \\
$\log_{10} E_{\rm K, iso}$ & $48,54.7$ & -- & $51.67^{+0.13}_{-0.12}$ \\
$\log_{10}\varepsilon_e$ & $-5,-0$ & $\varepsilon_e + \varepsilon_B < 1$ & $-0.52^{+0.03}_{-0.03}$ \\
$\log_{10}\varepsilon_B$ & $-5,-0$ & $\varepsilon_e + \varepsilon_B < 1$ & $-1.04^{+0.27}_{-0.23}$ \\
$t_{\rm jet}$ & $0,10^{\rm 5}$ & -- & $3.83^{+0.25}_{-0.22}$\\
$A_{\rm V, host}$ & $0,10$ & -- & $0.23^{+0.03}_{-0.03}$\\
\hline
\multicolumn{4}{c}{Model 2 - Off-axis and Energy Injection Allowed}\\
\hline
$p$ & $2.0 , 3.0$ & -- & $2.27\pm0.01$ \\
$\Gamma_0$ & $4-300$ & $0.0 \leq \gamma_{\rm collision}/\Gamma_0 \leq 0.5$ & $178^{+73}_{-58}$ \\
$\gamma_{\rm collision}$ & $2,50$ & $0.0 \leq \gamma_{\rm collision}/\Gamma_0 \leq 0.5$ & $44^{+4}_{-6}$ \\
$\theta_{\rm j}$ & $0.02,0.26$ & $0 \leq \theta_{\rm v} / \theta_{\rm j} \leq 3$ & $0.238\pm0.004$ \\
$\theta_{\rm v}$ & $\sin( 0, \pi/2)$ & $0 \leq \theta_{\rm v} / \theta_{\rm j} \leq 3$ & $0.011^{+0.006}_{-0.005}$ \\
$\log_{10} n_0$ & $-5,2$ & -- & $1.15\pm0.02$ \\
$\log_{10} E_{\rm K, iso, 0}$ & $48,55$ & -- & $50.57^{+0.05}_{-0.04}$ \\
$\zeta$ & $1,20$ & -- & $9.50^{+1.02}_{-1.13}$ \\
$s$ & $1,10$ & -- & $1.18^{+0.04}_{-0.03}$ \\
$\log_{10}\varepsilon_e$ & $-5,-0.05$ & $\varepsilon_e + \varepsilon_B < 0.9$ & $-0.43\pm0.01$ \\
$\log_{10}\varepsilon_B$ & $-5,-0.05$ & $\varepsilon_e + \varepsilon_B < 0.9$ & $-1.95\pm0.03$ \\
\hline
\multicolumn{4}{c}{Model 3 - Off-axis and Structured Jet Allowed}\\
\hline
$p$ & $2.0 , 3.0$ & -- & $2.33\pm 0.03$ \\
$\Gamma_0$ & $4-300$ & -- & $229^{+57}_{-56}$ \\
$\theta_{\rm c}$ & $0.02,0.26$ & $2 \leq \theta_{\rm j} / \theta_{\rm c} \leq 5$, $0 \leq \theta_{\rm v} / \theta_{\rm c} \leq 3$  & $0.086^{+0.016}_{-0.011}$ \\
$\theta_{\rm j}$ & $0.06,0.78$ & $2 \leq \theta_{\rm j} / \theta_{\rm c} \leq 5$ & $0.424^{+0.071}_{-0.052}$ \\
$\theta_{\rm v}$ & $\sin( 0, \pi/2)$ & $0 \leq \theta_{\rm v} / \theta_{\rm j} \leq 3$ & $0.155^{+0.003}_{-0.002}$ \\
$\log_{10} n_0$ & $-5,2$ & -- & $1.09\pm0.05$ \\
$\log_{10} E_{\rm K, iso}$ & $48,55$ & -- & $52.24^{+0.07}_{-0.06}$ \\
$a$ & $1,10$ & -- & $2.56^{+0.10}_{-0.13}$ \\
$s$ & $1,10$ & -- & $2.67^{+0.60}_{-0.28}$ \\
$\log_{10}\varepsilon_e$ & $-5,-0.05$ & $\varepsilon_e + \varepsilon_B < 0.9$, $0 < \varepsilon_e - \varepsilon_B < 1$ & $-0.56^{+0.01}_{-0.02}$ \\
$\log_{10}\varepsilon_B$ & $-5,-0.05$ & $\varepsilon_e + \varepsilon_B < 0.9$, $0 < \varepsilon_e - \varepsilon_B < 1$ & $-2.10\pm 0.08$ \\
\hline
\enddata
\tablenotetext{a}{For Model 1: The accelerated electron distribution index, $p$; ambient medium number density, $n_0~({\rm cm}^{-3})$; isotropic equivalent kinetic energy, $E_{\rm K, iso}$ (erg); the fraction of energy accelerating electrons, $\varepsilon_e$; the fraction of energy in magnetic fields, $\varepsilon_B$; jet break time, $t_{\rm jet}$ (d); extinction contribution from the host galaxy, $A_{\rm V}$ (mag). For Model 2: The accelerated electron distribution index, $p$; bulk Lorentz factor, $\Gamma_0$; Lorentz factor of decelerating blastwave at energy injection, $\gamma_{\rm collision}$; the jet half-opening angle, $\theta_{\rm j}$;  viewing angle, $\theta_{\rm v}$; ambient medium number density, $n_0~({\rm cm}^{-3})$; isotropic equivalent kinetic energy (initial), $E_{\rm K, iso, 0}$; fractional increase in total kinetic energy at end of injection, $\zeta$; stratification index for injected shell $E(>\Gamma)\geq \Gamma^{1-s}$, $s$; the fraction of energy accelerating electrons, $\varepsilon_{\rm e}$; the fraction of energy in magnetic fields, $\varepsilon_{\rm B}$. For Model 3: The accelerated electron distribution index, $p$; bulk Lorentz factor, $\Gamma_0$; the inner core of the jet, $\theta_{\rm c}$; the jet half-opening angle, $\theta_{\rm j}$; viewing angle, $\theta_{\rm v}$; ambient medium number density, $n_0~({\rm cm}^{-3})$; isotropic equivalent kinetic energy, $E_{\rm K, iso}$; the power-law index of the jet energy structure, $E_{\rm K, iso}(\theta_{\rm v}, \theta_{\rm c}) \propto (1+(\theta_{\rm v}/\theta_{\rm c})^2)^{-s/2}$, $s$; the power-law index of the jet velocity structure ($\Gamma_0(\theta_{\rm v}, \theta_{\rm c}) \propto (1+(\theta_{\rm v}/\theta_{\rm c})^2)^{-a/2}$, $a$; the fraction of energy accelerating electrons, $\varepsilon_{\rm e}$; the fraction of energy in magnetic fields, $\varepsilon_{\rm B}$.}
\tablenotetext{b}{Priors are uniform in the range shown except $\theta_{\rm v}$ which follows a sin distribution.}
\end{deluxetable*}

\section{Afterglow Modeling}
\label{sec:afterglowmodel}
\subsection{Model 1 - Forward Shock Only, On-axis Modeling}
\label{sec:GAMMA}

We first model the radio to X-ray afterglow emission of GRB\,250419A assuming a simple standard on-axis GRB forward shock model (``Model 1''). We utilize the modeling framework laid out in \citet{2014ApJ...781....1L}, which fits the data using a Markov Chain Monte Carlo (MCMC) method through the use of \texttt{emcee} \citep{2019JOSS....4.1864F} in order to determine $p$, $E_{\rm K, iso}$, $n_0$, $\varepsilon_{\rm e}$, $\varepsilon_{\rm B}$, and $t_{\rm jet}$. The model also fits for extinction along the line of sight due to the host galaxy ($A_{\rm V, host}$), and takes into consideration Milky Way dust extinction ($A_{\rm V, host} \approx 0.082~$mag). We further require $\varepsilon_{\rm e} + \varepsilon_{\rm B} < 1$ and assume the fraction of participating non-thermal electrons ($f_{\rm NT}$) is 1 \citep{2005ApJ...627..861E}. This framework also incoporates the scatter effects of scintillation \citep{1990ARA&A..28..561R, 2006ApJ...636..510G}, as well as the effects of inverse Compton cooling \citep{2001ApJ...548..787S, 2015ApJ...814....1L}  and Klein Nishina corrections \citep{2009ApJ...703..675N, 2021MNRAS.504..528J, 2024ApJ...970..135M}. 

\begin{figure*}
    \centering
    \includegraphics[width = 0.99\textwidth]{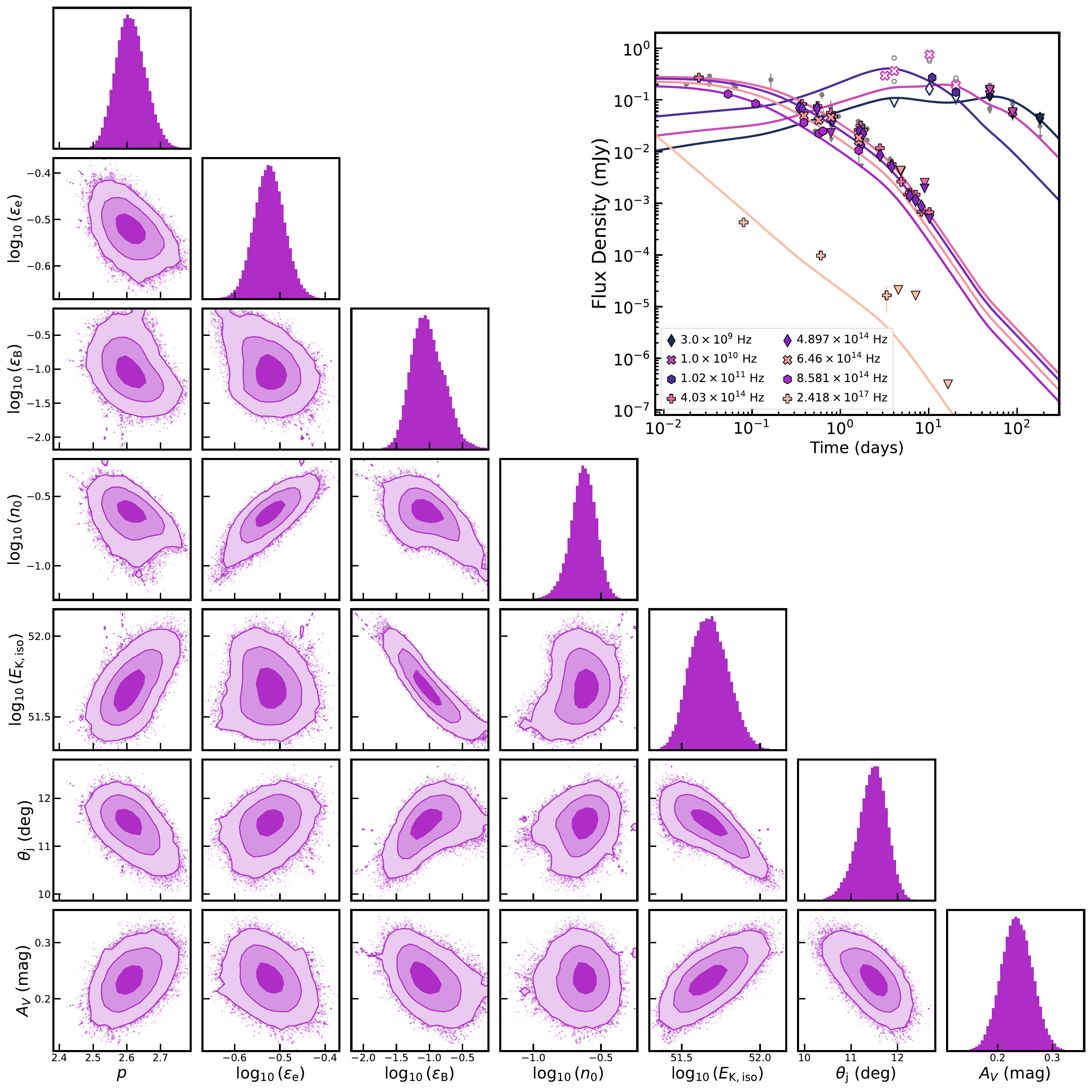}
    \caption{The model posterior samples for the on-axis forward shock fit (Model 1) of the afterglow of GRB\,250419A \citep{2014ApJ...781....1L}, as described in Section~\ref{sec:GAMMA}. {\it Top Panel:} The best fit model light curves compared to the afterglow data for a selection of observing frequencies (colored points), and the rest of the data (grey points). Open symbols represent masked data and triangles represent non-detections.}
    \label{fig:Gamma_results}
\end{figure*}

As determined in Section~\ref{sec:radio_afterglow}, the observed radio afterglow at $\delta t \lesssim 21~$d does not follow the expected forward shock behavior, and as a result we mask this data for this fit, though we leave the millimeter data unmasked in order to provide an anchor between the optical and X-ray data at early ($\delta t \lesssim 10~$d) times and the radio data at late times ($\delta t \gtrsim 10~$d). We further employ a $10\%$ uncertainty floor in order to prevent high signal-to-noise data from driving the afterglow fit. We fit the radio to X-ray afterglow of GRB\,250419A using 128 walkers and 5000 steps, and discard the initial $\sim 8\%$ steps as ``burn-in''. 
The best-fit model parameters are $p \approx 2.6$, $E_{\rm K, iso} \approx 5 \times 10^{51}~$erg, $n_0\approx0.2~{\rm cm}^{-3}$, $\varepsilon_{\rm e} \approx 0.3$, $\varepsilon_{\rm B} \approx 0.08$, and $t_{\rm jet}\approx3.9~$d, resulting in $\theta_{\rm j} \approx 0.2~{\rm rad}~(11^\circ)$, and $A_{\rm V, host}\approx 0.24~$mag, and is displayed in Figure~\ref{fig:Gamma_results}. The median and $68\%$ confidence intervals are stated in Table~\ref{tab:redback}, with the full posteriors displayed in Figure~\ref{fig:Gamma_results}. The break frequency ordering is $\nu_{\rm sa} < \nu_{\rm m} < \nu_{\rm c}$ for the majority of the detected afterglow (from $\delta t \approx 0.2$--$130~$d).

At $\delta t \approx 0.6~$d, this model results in $\nu_{\rm c} \approx 2 \times 10^{14}~$Hz, indicating that $\nu_{\rm c}< \nu_{\rm O} < \nu_{\rm X}$, inconsistent with our expectations of $\nu_{\rm O} < \nu_{\rm c}< \nu_{\rm X}$ (Section~\ref{sec:xray_optical}). This can be reconciled due to $A_{\rm V, host} \approx 0.24~$mag suppressing the optical afterglow by upto a factor of $\sim 2$ and resulting in a less steep measured $\beta_{\rm OX}$. While $t_{\rm jet}$ is somewhat later than our estimates of $\sim 2.0$--$2.9$, the resulting $\theta_{\rm j}$ is consistent with our expectations (Section~\ref{sec:xray_optical}). 

The best-fit light curve model is shown in Figure~\ref{fig:Gamma_results}. The model fits most of the data within a factor of $\sim 2$, and best reproduces the optical afterglow---which is not surprising, because the optical band is over-represented in the afterglow data compared to radio and X-ray observations. The model underpredicts the initial X-ray detection by a factor of $\sim 2$ and 
overpredicts the following X-ray detections by a factor of $\sim 2$--$4$. Given the limited X-ray data, it is difficult to determine whether the X-ray afterglow deviates from the forward shock model, or if rather the model is over-tuned to the optical data. 
For the majority of the radio afterglow, the model predicts the data within a factor of $\sim 2$, though it is typically underpredicted rather than over-predicted. The radio data at $\delta t \approx 4$--$11~$d is underpredicted by a factor of $\sim 2$--$5$, as expected due to its non-standard behavior (Section~\ref{sec:radio_afterglow}).

As mentioned in Section~\ref{sec:radio_afterglow}, in order to explain the radio afterglow, an alternative emission mechanism would need to be invoked, likely a reverse shock or energy injection \citep{1998ApJ...496L...1R,1998A&A...333L..87D,  2000ApJ...535L..33S, 2000ApJ...532..286K, 2001ApJ...552L..35Z, 2002ApJ...566..712Z, 2003ApJ...595..950Z, 2005ApJ...625..263G, 2005ApJ...628..315Z}. We explore the energy injection scenario in the following section, and briefly explore the implications of the reverse shock scenario here. A reverse shock naturally produces frequency dependent emission that typically dominates at radio frequencies \citep[e.g.][]{2000ApJ...535L..33S, 2005ApJ...625..263G}, which could allow for the optical and X-ray afterglow fit to remain unchanged while explaining the radio excess seen at $\delta t \lesssim 11~$days. Since the optical afterglow is already well described by Model 1, it is expected that any reverse shock emission would have marginal effects on the optical afterglow. With this, we can place constraints on the deceleration time of $t_{\rm dec} \lesssim 0.02~$d, the time of the first optical detection (Section~\ref{sec:optical_followup}), which would have the intended effect of not producing excess optical emission.

An initial attempt at a forward shock $+$ reverse shock model was made, but it was difficult to achieve an appropriately late radio bump with physically motivated parameters that also did not violate the early optical afterglow. For example, a reverse shock with $t_{\rm dec} \approx 75~$s, break frequencies of $\nu_{\rm a, rs} \approx 10^{13}~$Hz, $\nu_{\rm m, rs} \approx 10^{14}~$Hz, $\nu_{\rm c, rs} \approx 5\times 10^{15}~$Hz, peak flux of $F_{\nu, \rm m, rs} \approx 2 \times 10^{4}~$mJy, and $g \approx 5$ (where $g$ is related to the evolution of the Lorentz factor and radius of the ejecta such that $\Gamma \propto R^{-g}$) well describes the observed radio behavior without over-predicting the optical and X-ray. However, these parameters are not self-consistent with the forward shock model, and attempts to make the forward shock and reverse shock models consistent resulted in a model that violated the observed optical behavior. 

\subsection{Model 2 - Off-axis and Energy Injection Allowed}
\label{sec:redback_model2}

A limitation of the above afterglow model is that it requires the GRB to be observed on-axis ($\theta_{\rm v} = 0$). Given that GRB\,250419A is an XRF, we additionally model the radio to X-ray afterglow with a code that allows $\theta_{\rm v}$ to vary, in order to see if an off-axis model is preferred, and allow for energy injection as a refreshed shock (``Model 2'').

For Model 2 we utilise the {\sc redback} python package tophat jet afterglow model with a refreshed shock option, {\texttt tophat\_redback\_refreshed}\footnote{Redback version 1.15.1} \citep{lamb2018, lamb2020, sarin2024}. The optical data were corrected for extinction due to the Milky Way ($A_V = 0.08$ and $E(B-V) = 0.026$) and the host galaxy ($A_{V,{\rm host}} = 0.24$ and $E(B-V)_{{\rm host}}= 0.088$; see Model 1) prior to fitting.
The model parameters include the standard forward shock parameters in Model 1, e.g. $p$, $n_0$, $\varepsilon_{\rm e}$, and $\varepsilon_{\rm B}$, as well as the jet opening angle $\theta_{\rm j}$, the initial isotropic kinetic energy $E_{\rm K, iso,0}$, and the initial Lorentz factor $\Gamma_0$. Additionally, we allow $\theta_{\rm v}$ to vary, and incorporate energy injection under the standard refreshed shock model of slower moving ejecta catching up to decelerating higher-$\Gamma$ material. In this refreshed shock model framework, the free parameters are the Lorentz factor of the decelerating blast wave ($\gamma_{\rm collision}$), the fractional increase in total kinetic energy at end of injection ($\zeta$) and the stratification index for injected shell ($E_{\rm K, iso}(>\Gamma)\geq \Gamma^{1-s}$). We note, however, that this physical picture is not consistent with 3D hydrodynamic jet simulations \citep{2025ApJ...992L...3G}---so for our purposes we regard this model as generic energy injection which can arise from a variety of mechanisms (e.g. radial or angular stratification of the outflow). 

\begin{figure*}
    \centering
    \includegraphics[width = 0.99\textwidth]{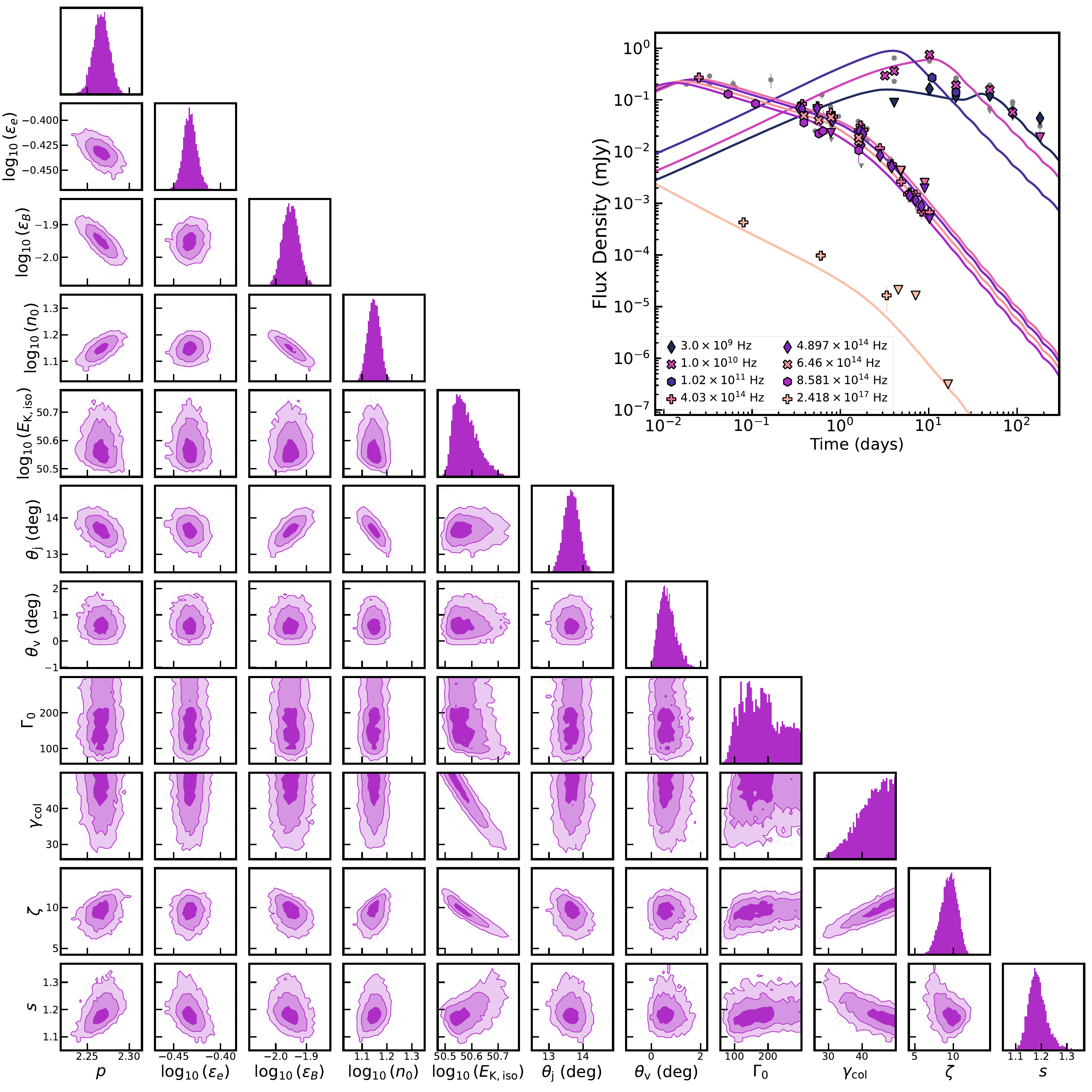}
    \caption{The model posterior samples for the \textsc{redback} fit of the afterglow of GRB\,250419A that allows for a jet viewed off-axis and energy injection, (Model 2) as described in Section~\ref{sec:redback_model2}. {\it Top Panel:} The best fit model light curves compared to the afterglow data for a selection of observing frequencies (colored points), and the rest of the data (grey points).}
    \label{fig:redback_results_model2}
\end{figure*}

The model priors are set to the ranges listed in Table \ref{tab:redback}, along with any enforced conditions.
The model and data are fit using a nested sampling algorithm, {\sc nessai}\footnote{Nessai version 0.15.2} \citep{williams2021}, which performs Bayesian inference while incorporating normalizing flows. The median and $68\%$ confidence intervals are stated in Table~\ref{tab:redback}, and the full posteriors along with the best fit light curve model are shown in Figure~\ref{fig:redback_results_model2}.

The median model parameters from Model 2 that can be directly compared to Model 1 (Section~\ref{sec:GAMMA}) are $p \approx 2.3$, $n_0 \approx 14~{\rm cm}^{-3}$, $\varepsilon_{\rm e} \approx 0.4$, and $\varepsilon_{\rm B} \approx 0.01$, and $\theta_{\rm j} \approx 0.24~$rad ($\sim 14^\circ$). Overall, these values are similar to those found in Model 1, while some differences, such as $\varepsilon_{\rm B}$ and $n_0$, may be due to degeneracies in fitting the parameters, as seen in Figure~\ref{fig:redback_results_model2}. We discuss some of the parameters of interest below.

For the model, the prior range and conditions were chosen so that the energy injection can be passive, i.e., no noticeable effect on the dynamics or afterglow light curve.
In other words, additional energy injection is not required if the forward shock alone is sufficient to describe the multiwavelength data.
However, from the posterior values, it is clear that some level of energy injection is preferred by the data when fit with this model.
The model fit is consistent with the majority of the data, although the model slightly underpredicts the X-ray data.
A jet break at $\sim 2$ days is required (similar to expectations from Section~\ref{sec:xray_optical}), with the pre-break optical data being mostly well reproduced by the model.
However, post-break the optical decline seen in the data is marginally faster than that of the model (we discuss this more below).

\begin{figure*}
    \centering
    \includegraphics[width = 0.98\textwidth]{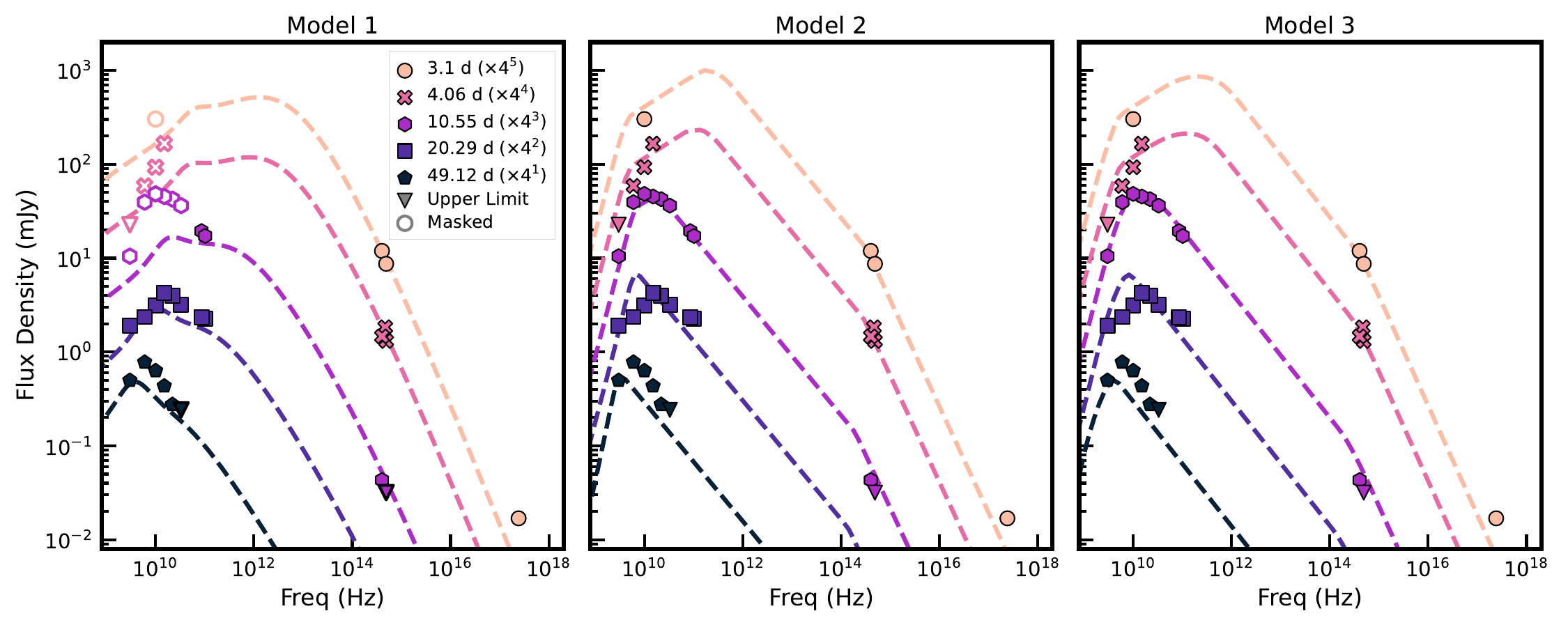}
    \caption{The radio to X-ray spectral energy distributions (SEDs) of GRB\,250419A at selected times. Triangles represent upper limits. The panels show the model SEDs for a standard on-axis synchrotron afterglow (``Model 1'', Left, Section~\ref{sec:GAMMA}), a synchrotron afterglow model that allows for energy injection and an off-axis viewing angle (``Model 2'', Middle, Section~\ref{sec:redback_model2}), and a synchrotron afterglow model that allows for a structured jet and an off-axis viewing angle (``Model 3'', Right, Section~\ref{sec:redback_model3}). Empty symbols indicate data that were masked for the afterglow model fitting.}
    \label{fig:SED_model_comparison}
\end{figure*}

At radio wavelengths, the \texttt{tophat\_redback\_refreshed} model used includes synchrotron self-absorption, the effects of which can be seen at 3 and 10 GHz light curves.
For the radio data that is below $\nu_{\rm m}$, the jet break results in a shallower incline or a flattening of the light curve where the emission is self-absorbed.
Post-break, the radio data will peak when $\nu_{\rm m}$ passes the observation band, and this can be seen in the data as increasingly later peaks for lower frequencies.
Where the emission is self-absorbed, when $\nu_{\rm sa}$ passes the observation band, we will see a rising light curve until the passage of $\nu_{\rm m}$, this effect is best seen for the 3~GHz light curve.
This variability recreates the observed radio wavelength data, noting that scintillation effects are not included in this model. As a result of this model, the radio SED at $\delta t \approx 10.55~$d is well fit (Figure~\ref{fig:SED_model_comparison}), and the model results in $\nu_{\rm m} \approx 15~$GHz, removing the need for a reverse shock to explain the radio data. However, this results in the radio data at $\delta t \gtrsim 10.55~$d to be mostly underpredicted.

We note that the model indicates that energy injection is required to commence at early times (at $\sim 1$ hour in the observer frame), and the order unity index, $s\approx1.18$ will make the temporal decline shallower than that expected via standard closure relations while energy injection persists (e.g. Section~\ref{sec:xray_optical} and \ref{sec:opt_comparison}). Additionally, the order of unity $s\approx 1.18$ indicates that fractional energy increase of $\zeta \approx 9.5$ is never realized, instead the energy only increases by a factor of $\sim 1.8$ over the course of our observed data.

One difference between Model 2 and Model 1 (Section~\ref{sec:GAMMA}) is the preferred value for the distribution of accelerated electrons, where Model 1 returns $p\sim2.6$ and Model 2 returns $p\sim 2.3$.
This lower value of $p$ in  Model 2 appears to be driven by $\nu_{\rm c}$ passing through the optical band around $\delta t \approx 1~$day. As a result, $p$ must be $\sim 2.3$ to reconcile the $\beta_{\rm OX} \approx -1$ (not corrected for extinction, Figure~\ref{fig:XrayOptical}). A higher value of $p$ would lead to an even more apparent 
underprediction of the X-ray afterglow, which occurs in Model 1.

\begin{figure*}
    \centering
    \includegraphics[width = 0.99\textwidth]{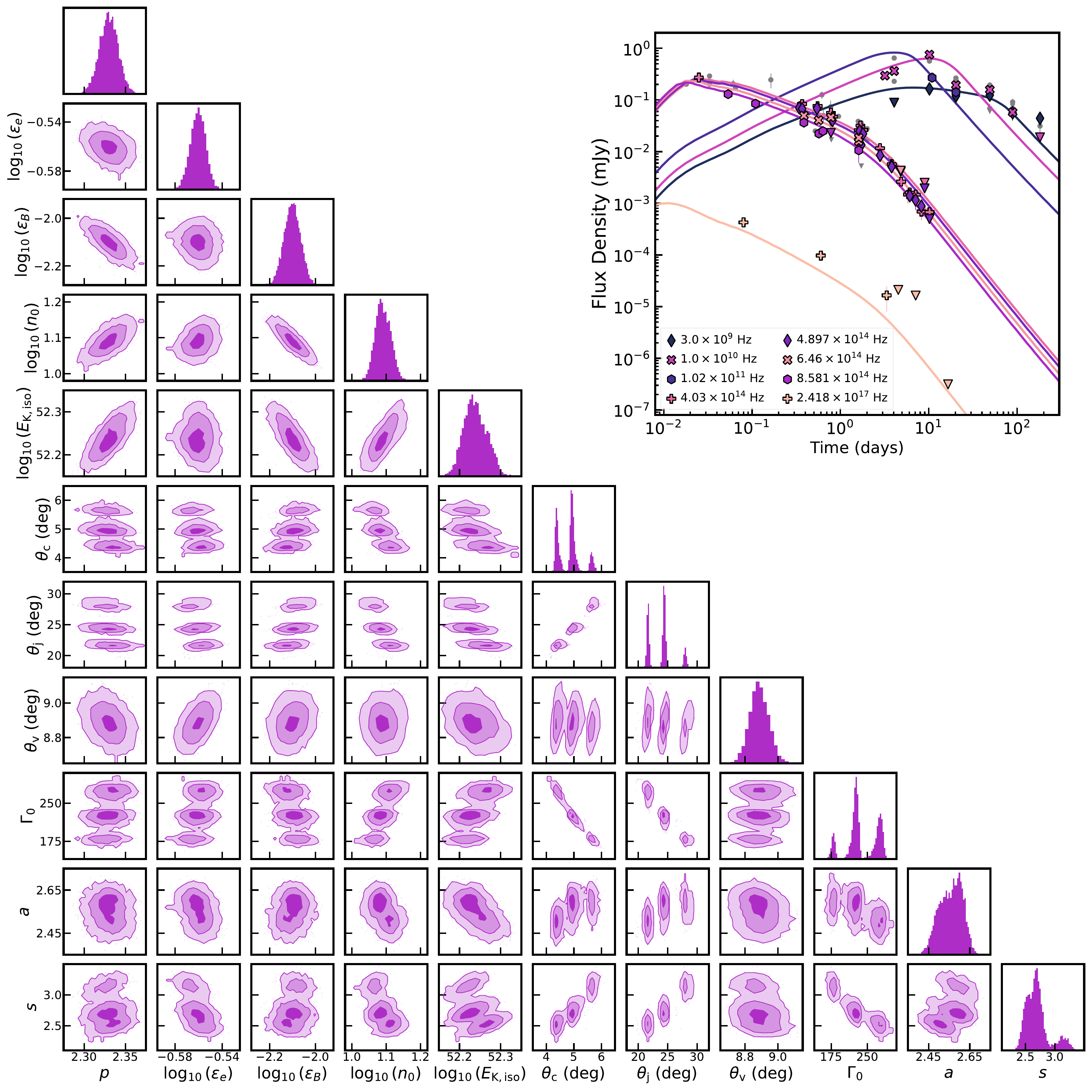}
    \caption{The model posterior samples for the \textsc{redback} fit of the afterglow of GRB\,250419A that allows for a structured jet viewed off-axis (Model 3), as described in Section~\ref{sec:redback_model3}. {\it Top Panel:} The best fit model light curves compared to the afterglow data for a selection of observing frequencies (colored points), and the rest of the data (grey points).}
    \label{fig:redback_results_model3}
\end{figure*}

The prior for the initial Lorentz factor has a broad range, $4<\Gamma_0<300$, where values below $\sim30$ are expected to result in jets that are opaque to gamma-rays.
The peak time for lower Lorentz factor jets is later for ``on-axis'' observers, as the peak time is $\propto \Gamma_0^{-8/3}$ -- the same arguments for the ``off-axis'' timing and peak flux hold as for a regular GRB Lorentz factor jet.
A lower Lorentz factor would indicate a preference for a `dirty-fireball' scenario, where gamma-rays are absorbed before the outflow becomes optically thin.
However, the posterior distribution indicates a model-data preference for $\Gamma_0=178^{+73}_{-58}$, a typical GRB jet Lorentz factor.

Overall, Model 2 does a better job at reproducing the observed data than Model 1, which is to be expected given that Model 2 has more free parameters. The invocation of energy injection can reproduce the shallowness of the optical decline, $\alpha\approx-0.5$ (Figure~\ref{fig:XrayOptical}), however physically this only indicates that energy is arriving at later times, and could have multiple physical explanations. For example, a mildly off-axis or jet edge orientation can be a natural explanation for the continuous injection of energy \citep[e.g.][]{2006ApJ...641L...5E, 2011ApJ...727..132M}.

\subsection{Model 3 - Off-axis and Structured Jet Allowed}
\label{sec:redback_model3}

One physical interpretation of the energy injection in Model 2 is angular structure. Here we explicitly fit the data to a structured jet model \citep[e.g.,][]{2021MNRAS.500.3511G}. We use the power law structure in \texttt{alternativepowerlaw\_redback} (``Model 3''), in which the jet is assumed to have an inner core ($\theta_{\rm c}$) and outer jet ($\theta_{\rm j}$), with an assumed structured jet energy ($E_{\rm K, iso}(\theta_{\rm v}, \theta_{\rm c}) \propto (1+(\theta_{\rm v}/\theta_{\rm c})^2)^{-s/2}$) and jet velocity ($\Gamma_0(\theta_{\rm v}, \theta_{\rm c}) \propto (1+(\theta_{\rm v}/\theta_{\rm c})^2)^{-a/2}$). This model allows for $\theta_{\rm v}$ to vary in the same way as Model 2 (Table~\ref{tab:redback}), and the other free parameters are $p$, $\Gamma_0$, $n_0$, $E_{\rm K,iso}$, $\varepsilon_{\rm e}$, $\varepsilon_{\rm B}$. The posterior samples are presented in Figure~\ref{fig:redback_results_model3} and the median and $68\%$ confidence intervals are presented in Table~\ref{tab:redback}. 

This fit reproduces the observed afterglow similarly well to Model 2.
Overall, the fit parameters of $p$, $n_0$, $\varepsilon_{\rm e}$ and $\varepsilon_{\rm B}$ are similar to Model 2. 
We find $\theta_{\rm c} \approx 4$--$6^\circ$, $\theta_{\rm j} \approx 20$--$30^\circ$, and $\Gamma_0 \gtrsim 150$. The viewing angle is constrained to $\theta_{\rm v} \approx 9^\circ$. In other words, this model indicates that the afterglow of GRB\,250419A is consistent with a structured relativistic jet viewed moderately off-axis.

\section{Discussion}
\label{sec:Discussion}

\subsection{Implications of Prompt Emission and Afterglow Models}

We next explore how the models presented above affect our interpretation of the progenitor of GRB\,250419A.
Models 1 and 2 find $\theta_{\rm j} \approx 10$--$14^\circ$, similar to our expectations from jet break arguments (Section~\ref{sec:xray_optical}), whereas Model 3 finds a narrower jet core of $\theta_{\rm c} \approx 4$--$6^\circ$ and a wider jet of $\theta_{\rm j} \approx 20$--$30^\circ$.
From the Model 1 and Model 3 parameters we calculate the total beaming corrected kinetic energy to be $E_{\rm K} \approx 10^{50}~$erg, whereas Model 2 is an order of magnitude lower at $E_{\rm K} \approx 10^{49}~$erg.
These values are on the low end of the distribution for traditional long GRBs \citep[e.g.][]{2001ApJ...562L..55F, 2002ApJ...571..779P, 2003ApJ...594..674B,2014PASA...31....8G,2018IJMPD..2742002V, 2021ApJ...911...14K}, indicating that GRB\,250419A may be an intrinsically low energy event. We derive the GRB efficiency for both models by calculating $\eta = E_{\rm iso, bol}/(E_{\rm iso, bol}+E_{\rm K, iso, total})$ \citep{2004ApJ...613..477L}, and find $\eta_{\rm bb} \approx 0.1$--$1$ and $\eta_{\rm pl} \approx 0.03$--$1$. These values are consistent with both the observationally derived distribution, as well as theoretical expectations \citep{2001ApJ...551..934K, 2004ApJ...613..477L, 2007ApJ...655..989Z, 2015ApJS..219....9W, 2016MNRAS.461...51B, 2018ApJS..236...26L, 2023ApJ...944L..57L, 2024ApJ...972..195L}, though we note the range is very broad. 

Our afterglow modeling cannot distinguish between an on-axis burst and an off-axis structured jet, however the low kinetic energy seems to be ubiquitous.
The initial Lorentz factor is estimated to be $\Gamma_0 \gtrsim 100$ through a variety of methods, suggesting the jet that in GRB\,250419A was ultrarelativistic and not a dirty fireball. 

In addition to describing the observed afterglow well, the invocation of an energy injection event in Model 2 and the structured off-axis jet in Model 3 may be observationally supported by the prompt emission. As mentioned in Section~\ref{sec:svom_detection}, the initial prompt emission was proceeded by a precursor pulse $\sim 120~$s prior to the trigger. An analysis on the XRFs and GRBs detected by BeppoSAX found that $\sim 10~\%$ of events also exhibited precursor X-ray emission $\sim 10$--$100~$s prior to trigger \citep{2026arXiv260600225P}, though the precursor emission was typically softer, which is not the case for GRB\,250419A.

While the black body (thermal) model of the prompt emission may support a shock breakout scenario to explain the precursor, this scenario is not favored by the afterglow. We can additionally reject the shock breakout scenario as its expected that the emission should reach $\gtrsim 50 \Gamma_0~$keV \citep[e.g.][]{{2010ApJ...725...63B, 2012ApJ...747...88N}}, much higher than the observed $\sim 2.3~$keV (Section~\ref{sec:svom_detection}).
Instead, a weak or intermittent jet \citep{2007ApJ...670.1247W, 2016ApJ...826..180L, 2016ApJ...833..116G}, or even a wobbling jet coming in and out of the line of sight \citep{2025ApJ...992L...3G}, would deposit additional energy into the surrounding medium, and is a natural explanation for the energy injection invoked in Model 2 or the off-axis structured jet invoked in Model 3. Furthermore, in Model 3 the line-of-sight Lorentz factor is only $\sim 35$, which could be an explanation for the low peak energy of the prompt emission.

\subsection{GRB\,250419A in the context of XRFs}

Here we explore how GRB\,250419A fits into the current sample of both pre-EP/SVOM XRFs \citep[e.g. XRFs\,020903, 030723, 060218, 080109, 100316D,][]{2004ApJ...606..994S, 2004ApJ...605..300H, 2006Natur.442.1014S, 2006Natur.442.1008C, 2006Natur.442.1014S, 2008Natur.453..469S, 2013ApJ...778...18M} and newly detected XRFs by EP/SVOM \citep[e.g. EP\,240414a, EP240801a/XRF 240801B, EP\,241113a, XRF\,241001A, EP\,250108a, EP\,250827a, EP\,260321a, ][]{2025NatAs...9.1073S, 2025ApJ...985...21Z, 2025arXiv250417034L, 2025arXiv251210239S, 2026arXiv260326213D, 2025ApJ...988L..34J, 2026arXiv260610011R, 2026arXiv260420346S} with well studied broad-band emission. 

Among the known and well-studied population of XRFs, we identify two emerging populations: those that are ``afterglow-bright'', and those that are ``afterglow-faint''. The afterglow-bright population is characterized by long-lasting ($> 1~$d), fading (typically power-law like fading), broad-band (X-ray to radio), and luminous emission (band specific but comparable to GRB afterglows). In contrast, the afterglow-faint XRFs typically do not have detectable X-ray emission after $> 1~$d, have optical light curves dominated by a supernova rather than a non-thermal power-law decay, and have either sub-luminous or undetected radio emission. Within the afterglow-faint population is XRF\,060218, XRF\,080109, XRF\,100316D, EP250108a, EP250827b, and EP260321a, and these events typically have prompt X-ray emission lasting $\sim 100$--$1000~$s \citep{2006Natur.442.1008C, 2008Natur.453..469S, 2011MNRAS.411.2792S, 2025arXiv250417034L, 2025arXiv251210239S, 2026GCN.44075....1H}. Within the afterglow-bright XRF population are XRF\,020903, XRF\,030723, EP\,240414a, EP240801a/XRF 240801B, EP\,241113a, and XRF\,241001A, and these events typically have prompt emission lasting $\sim 1$--$100~$s \citep{2004ApJ...602..875S, 2003GCN..2313....1P, 2025NatAs...9.1073S, 2025ApJ...988L..34J, 2025ApJ...986L...4H, 2026arXiv260326213D, 2026arXiv260420346S}.
One explanation for the afterglow-faint population may be shock breakout events, which could explain the longer lasting prompt emission \citep[e.g. ][]{2012ApJ...747...88N}. 

As discussed in depth in this paper, GRB\,250419A belongs to the afterglow-bright population, and as a result we will focus the rest of this discussion on that category. The broad-band emission of this subset of XRFs has been modeled within the afterglow framework, and the results have indicated that some may originate from wide, quasi-spherical explosions, some may be the result of GRBs observed moderately off-axis \citep{2004ApJ...605..300H, 2025ApJ...985...21Z}, and some may in fact be low-$\Gamma$ dirty fireballs \citep{2026arXiv260326213D}. Thus far, the evidence supporting any of these claims is not irrefutable, and likely more events or more thorough modeling of the existing events as a population will be needed to substantiate these claims. Interestingly, one common thread seems to be true for the vast majority of the afterglow-bright XRFs: the total beaming corrected energy of the explosion is $\lesssim {\rm few}\times10^{50}~$erg \citep[][]{ 2004ApJ...605..300H, 2006ApJ...638..930S, 2013ApJ...778...18M,2025NatAs...9.1073S, 2026arXiv260420346S, 2025ApJ...988L..34J} (the one exception is EP241113a with $\sim 10^{51}~$erg, \citealt{2026arXiv260326213D}). Furthermore, some of these events are consistent with $\Gamma \gtrsim 100~$ and $\theta_{\rm v} < \theta_{\rm j}$, indicating that at least some XRFs originate from on-axis ultrarelativistic GRBs with low intrinsic energies. Based on our afterglow modeling of GRB\,250419A, it appears to be within the ultrarelativistic and low energy group of XRFs, though it may be observed off-axis.

\subsection{Future Constraints on Dirty Fireballs and Off-axis GRBs}

Here we explore the future of detecting dirty fireballs and off-axis GRBs. In this work, we attempted to identify an off-axis GRB or dirty fireball through the follow-up of a high energy event that appeared to be an outlier on the Amati diagram. Through multiwavelength modeling, we determined that the event was consistent with being an off-axis structured jet and unlikely to be a dirty fireball. However, we could not distinguish a marginally off-axis jet from an on-axis jet with other sources of energy injection (e.g., a radially stratified outflow). 

The advantage of identifying dirty fireballs and off-axis GRBs through the detection of XRFs is twofold: the precise explosion time is known, and the prompt emission energy is constrained, both of which are difficult to determine with optically discovered transients. Therefore, continued multiwavelength follow-up of XRFs, in addition to robust modeling of the population and further simulations of the expected emission, is still the best way to understand these events. 

Our analysis of GRB\,250419A could have benefited from earlier radio observations that occurred prior to the observed jet break and concurrent with the X-ray and optical shallow decay. Such radio observations not only would have better constrained $\Gamma_0$ through scintillation and temperature arguments, but also would have better anchored the location of the break frequencies and any alternative emission features. To this end, rapid response observing modes on radio telescopes, such as what is implemented on the Australia Telescope Compact Array \citep{2021MNRAS.503.4372A, 2026ApJ...997L...1L} and the Arcminute Microkelvin Imager - Large Array \citep{2018MNRAS.473.1512A, 2026arXiv260813697R} and what is proposed for the Square Kilometer Array \citep{2026arXiv260703024A}, will help better enable us to understand the broad-band emission from XRFs and their connection to off-axis GRBs and dirty fireballs. 

Capturing the rising optical emission would have also helped break degeneracies in the observing angle and initial Lorentz factor. Serendipitous detections by high-cadence optical observations have already provided a proof of concept for a handful of events, although interpretation (dirty fireball vs. off-axis GRB) has been hindered by a lack of detected prompt emission \citep{2025MNRAS.537.2362P, 2026arXiv260817242R, 2025ApJ...985..124L}. 
The Argus Array \citep{2022PASP..134c5003L} is one promising future facility that will be able to sample the rise of a large number of afterglows from X-ray flashes. 

\section{Conclusion}
\label{sec:Conclusion}

In this work, we presented observations and analysis of the multiwavelength afterglow of GRB\,250419A, an XRF discovered by SVOM. We have come to the following conclusions:
\begin{itemize}
    \item The prompt emission of GRB\,250419A indicates it is an XRF and places it in an undersampled portion of the Amati diagram. While it appears to be an outlier in the Amati diagram, additional XRFs (detected by e.g. Einstein Probe and SVOM) populating the parameter space are required to make statistical claims on the uniqueness of GRB\,250419A.
    \item Based on luminosity and prompt energy arguments, the X-ray, optical, and radio afterglow of GRB\,250419A is somewhat underluminous but mostly consistent with the overall long GRB population. \item The optical afterglow in particular is characterized by a shallow decay phase at $\lesssim 2~$d that appears to be inconsistent with expectations for the forward shock afterglow. Similarly, the radio afterglow demonstrates deviations from forward shock expectations at $\sim 10.6~$d. 
    \item The broadband afterglow of GRB\,250419A is reasonably fit by a low energy ($\sim 10^{49}$--$10^{50}~$erg) relativistic ($\Gamma_0 \gtrsim 100$) jet, indicating that the event was not a low-$\Gamma_0$ dirty fireball. The shallow early decline in the optical and X-ray afterglow can be explained by either an off-axis structured jet or an on-axis refreshed shock, but the data do not conclusively confirm either model.
\end{itemize}

GRB\,250419A adds to the increasing population of afterglow-bright XRFs, and indicates at least some XRFs are long GRBs at the lower end of the energy distribution.
As X-ray monitors such as Einstein Probe and SVOM continue to detect XRFs, it will become clearer how these events relate to traditional long GRBs in terms of both their prompt and broad-band properties. Future observations of XRFs with rapid response radio and high cadence optical observations will also provide valuable information on the early behavior and help constrain the initial Lorentz factor as well as viewing angle.


\begin{acknowledgments}

G.S. thanks Brian Metzger for the helpful discussions regarding prompt emission expectations.

G.S. is a Canadian SKA Scientist and is funded by the Government of Canada/est financé par le gouvernement du Canada.
G.S. and A.Y.Q.H. acknowledge support from a Sloan Research Fellowship (Award Number FG-2024-21320) from the Alfred P. Sloan Foundation, and a Scialog Award from the Research Corporation for Science Advancement. 

This work made use of data supplied by the UK Swift Science Data Centre at the University of Leicester. This research has made use of data obtained from the Chandra Data Archive provided by the Chandra X-ray Center (CXC). The National Radio Astronomy Observatory are facilities of the U.S. National Science Foundation operated under cooperative agreement by Associated Universities, Inc. This work is based on observations carried out under project numbers E24AN001 and E24AN002 with the IRAM NOEMA Interferometer. IRAM is supported by INSU/CNRS (France), MPG (Germany) and IGN (Spain). The Liverpool Telescope is operated on the island of La Palma by Liverpool John Moores University in the Spanish Observatorio del Roque de los Muchachos of the Instituto de Astrofisica de Canarias with financial support from the UK Science and Technology Facilities Council.

\end{acknowledgments}





%
\facilities{VLA, IRAM:NOEMA, Swift, CXO, Liverpool:2m, LDT, SOAR}

\software{CASA \citep{2007ASPC..376..127M}, pwkit \citep{2017ascl.soft04001W}, matplotlib \citep{matplotlib}, pandas \citep{pandas}, numpy \citep{numpy}, redback \citep{redback}, nessai \citep{nessai}
          }





\bibliography{library,journals_apj}
\bibliographystyle{aasjournalv7}



\end{document}